\documentstyle[preprint,aps]{revtex}
\tighten
\begin{document}
\draft
\preprint{WUGRAV 95-16}
\title{Statistics of the Microwave Background Anisotropies Caused\\
by Cosmological Perturbations of Quantum-Mechanical Origin}
\author{L. P. Grishchuk\footnote{grishchu@howdy.wustl.edu}}
\address{McDonnell Center for the Space Sciences, Physics Department\\
Washington University, St. Louis, Missouri 63130\\
and\\
Sternberg Astronomical Institute, Moscow University\\
119899 Moscow, V234, Russia}
\maketitle
\begin{abstract}
The genuine quantum gravity effects can already be around us.
It is likely that the observed large-angular-scale anisotropies
in the microwave background radiation are induced by
cosmological perturbations of quantum-mechanical origin.
Such perturbations are placed in squeezed vacuum quantum
states and, hence, are characterized by large variances of
their amplitude. The statistical properties of the anisotropies
should reflect the underlying statistics of the squeezed vacuum
quantum states. In this paper, the theoretical variances for the
temperature angular correlation function are described in detail.
It is shown that they are indeed large and must be present in the
observational data, if the anisotropies are truly caused by
the perturbations of quantum-mechanical origin. Unfortunately, these
large theoretical statistical uncertainties will make the extraction
of cosmological information from the measured anisotropies a much
more difficult problem than we wanted it to be.
This contribution to the Proceedings is largely based on
references~[42,8]. The Appendix contains an analysis of
the ``standard'' inflationary formula for density perturbations.
\end{abstract}
\newpage
\section{Introduction}
In the context of cosmology, the quantum gravitational physics is usually
understood as the early Universe physics at the Planck scale.
It is assumed that the gravitational field is fully quantized and the
Universe itself is, in a sense, quantized. (For a comprehensive recent review
of the structural issues in quantum gravity see [1]). However, there
exists also another
meaning in which we can speak of quantum gravity effects. To clarify the
difference, let us consider an analogy from condensed matter physics.
Imagine a crystal and various quantized excitations in it: phonons, rotons,
excitons, etc.. The creation and annihilation operators that the condensed
matter theorists write do not create and annihilate the crystal, they create
and annihilate excitations in the crystal. The excitations should not
necessarily be linear, one can take account of the higher-order corrections
too. The theory of the crystal excitations
is a fully legitimate quantum theory which does not attempt to quantize
``everything in sight''. Similarly, in our study below, we do not write the
creation and annihilation operators that create and destroy universes. Our
operators create and destroy perturbations in the Universe. Nevertheless,
the effects that we are studying are genuine quantum gravity effects in the
sense that they inherently contain all the three fundamental constants. The
gravitational constant and velocity of light enter because we deal with a
gravitational field (its energy-momentum tensor), the Planck constant enters
because we normalize the vacuum energy of the field to have ``a half of
the quantum'' in each mode.  All three fundamental constants combine in
the Planck length $l_{Pl}$, and this quantity naturally appears as
the coefficient in the most of our formulas. The sending of the $l_{Pl}$
to zero would eliminate the entire expression.

Now, why at all do we think that the anisotropies in the microwave
background radiation may have something to do with gravitational
quantum physics? Why, in the first place, is there such
a considerable interest to the measured large-angular-scale
anisotropies~[2]? After all, we have always knew that the observed
part of the Universe is homogeneous and isotropic only approximately,
on average, and we are certainly aware of large deviations on smaller
scales.

The point is that the photons of the microwave background have been
traveling to us for almost all duration of that characteristic time
which we call the age of the Universe and which is determined by
the value of the measurable quantity~---~the Hubble parameter.
The anisotropies on largest angular scales are specific in that
they are produced by cosmological perturbations on largest spatial scales.
In our context, the largest spatial scale means definitely larger than
any directly studied distance, and comparable or larger than the
characteristic cosmological distance~---~the Hubble radius~---~associated
with the Hubble parameter. In terms of the crystal analogy,
and assuming that the size of the crystal varies at some time scale,
we are dealing with excitations whose wavelengths are comparable
and longer than the light travel spatial scale associated with the
time variation.

The next question is whether we will be attempting to explain the
origin and nature of such long-wavelength cosmological perturbations or
we will be happy to simply accept their existence. There is no
logical contradiction in the second position. One can study the
perturbations with whatever accessible observational accuracy one has,
and then extrapolate back in time their evolution according to classical
dynamical laws. Typically, we will end up with a very anisotropic and
inhomogeneous universe at some very early times.
As an explanation, we will have nothing more to say except
that this was the Universe that was given to us for our future
life and study. A more appealing position, at least aesthetically, is to
try and find out a universal and reliable mechanism which could be capable of
generating the required perurbations in the originally homogeneous and
isotropic
Universe. The problem is not to increase the wavelengths of the perturbations
up to the size of the present Hubble radius (in the expanding Universe, the
wavelength is always growing with time) but to generate them.
In principle, one can imagine that such a mechanism may
have operated in the relatively recent Universe and may have not be related
to quantum gravity. In practice, taking into account everything
what we know about cosmology and the necessity to produce perturbations
with extremely long wavelengths, it is difficult to suggest such a
mechanism, especially if we do not want to make many additional hypotheses.
It appears that we can only rely on the quantum processes in the very early
Universe. The parametric interaction of quantized cosmological
perturbations with strong variable gravitational field of the very
early Universe provides us with such a possibility. The perturbations are
generated as a result of amplification of their zero-point quantum
oscillations. Returning to the laboratory physics analogy, one can recall that
a cavity filled with a dielectric medium and initially free of
electromagnetic radiation will eventually contain the radiation if the
parameters of the dielectric medium vary properly in time.
The quantum-mechanical (parametric)
mechanism of generating cosmological perturbations relies only on the
validity of the general relativity and the basic
principles of quantum field theory. The observational consequences
of this phenomenon we will study below.

The line of reasoning in this study can be summarized as follows.

We see the anisotropies in the microwave background radiation at
the largest angular scales~[2]. Observers convincingly argue that
this is a genuine cosmological effect.

If the large-angular-scale anisotropy in the microwave background is
really produced by cosmological perturbations (density perturbations,
rotational perturbations, gravitational waves), then their today's
wavelengths are of the order and longer than today's Hubble
radius $l_H$. Strictly speaking, all wavelengths give contributions
to the anisotropy at every given angular scale. But if the spectrum
of the perturbations is not excessively ``red'' or ``blue'', the dominant
contribution is provided by wavelengths indicated above. For instance,
the major contribution to the quadrupole anisotropy is provided
by wavelengths somewhat longer than $l_H$.

In the expanding Universe, the wavelengths of perturbations increase
in proportion to the cosmological scale factor. The wavelengths
that are longer than some length scale today have always been
longer than that scale in the past. Moreover, the wavelengths
of the perturbations of our interest are much longer than
the Hubble radius defined at the previous times, when one
goes back in time up to the era of primordial
nucleosynthesis --- the earliest era of which we have
observational data. It is hard to imagine (although it
does not seem to be logically impossible) that cosmological
perturbations of our interest, with such long wavelengths,
could have been generated by local physical processes during
the interval of time between the era of primordial nucleosynthesis
and now. We are bound to conclude that these perturbations were
generated in the very early Universe, before the era of primordial
nucleosynthesis. There is still 80 orders of magnitude, in terms
of energy density, to go from the era of primordial nucleosynthesis
to the Planck era; a lot of things could have happened in between.

The law of evolution of the very early Universe is not known,
but it is likely that it could have been significantly different
from the law of expansion of the radiation-dominated Universe.
If so, some amount of cosmological perturbations must have been
generated quantum-mechanically, as a result of parametric
interaction of the quantized perturbations with strong variable
gravitational field of the very early Universe. Gravitational waves
have been generated inevitably, while density and rotational
perturbations --- if we were lucky; see~[3] and references
therein. (If the cosmological scale factor has always been
the one of the radiation-dominated Universe, we must stop here,
because the parametric coupling vanishes in this case,
and cosmological perturbations cannot be amplified classically
and cannot be generated quantum-mechanically.) The amount
and spectrum of the generated perturbations depend on the law
of evolution of the very early Universe (the strength and variability
of the gravitational pump field), and this is how we can learn
about what was going on there. In particular, the law of evolution
of the very early Universe could have been of inflationary type.

If the cosmological perturbations were generated quantum-mechanically,
they should be placed in the squeezed vacuum quantum states~[4]
(for an introduction to squeezed states see, for example, Ref.~[5,6]
and the pioneering works quoted there). Squeezing of cosmological
perturbations might have degraded by now at short wavelengths
but should survive at long wavelengths, especially in the case
of gravitational waves.

The squeezed vacuum quantum states can only be squeezed
in the variances of phase which unavoidably means the increased
variances in amplitude. The statistical properties of the squeezed
vacuum quantum states are significantly different from the statistical
properties of the ``most classical'' quantum states --- coherent states.
This is well illustrated by the fact that the variance of the number
of quanta in a strongly squeezed vacuum quantum state is much larger
than the variance of the number of quanta in the coherent state with
the same mean number of quanta
$\langle N\rangle$, $\langle N\rangle \gg 1$.
For a squeezed vacuum state the variance is
$\langle N^2\rangle -\langle N\rangle^2
=2\langle N\rangle (\langle N\rangle +1)\gg \langle N\rangle$,
while for a coherent state it is
$\langle N^2\rangle -\langle N\rangle^2 = \langle N\rangle$.
In cosmology, the mean number $\langle N\rangle$ is a characteristic
of the expected mean square amplitude of cosmological perturbations,
while the variance $\langle N^2\rangle -\langle N\rangle^2$
is a characteristic of theoretical uncertainties in the amplitude.
These two characteristics are independent properties of a quantum state
or a stochastic process. Theoretical models may agree on $\langle N\rangle$
and disagree on variance or agree on variance and disagree
on $\langle N\rangle$.

The statistical properties of squeezed cosmological perturbations
will inevitably be reflected in statistical properties of the
microwave background anisotropies caused by them. Squeezing is
a phase-sensitive phenomenon, and to fully extract its properties
the quantum optics experimenters use the phase-sensitive detecting
techniques based on a local oscillator. In cosmology, we are very
far from being able to build a local oscillator, except of maybe,
in the distant future, for short gravitational waves.
Besides, in our study of the microwave background anisotropies,
we are interested in so long-wavelength perturbations that it
would take billions of years to wait for seeing the time dependent
oscillations of variances in the quadrature components of the
perturbation field. On the other hand, the amount of cosmological
squeezing is enormously greater than what is achieved in quantum
optics laboratory experiments. In cosmology, we can only rely
on the phase-insensitive, direct detection.  One can expect
that the underlying large variances of the amplitude of cosmological
perturbations should result in large statistical deviations from
the mean values for the microwave anisotropies.

A detailed study and proof of this statement is the purpose of this
work.

At this point it is necessarry to comment on the possible numerical
values of $\langle N\rangle$ (and, hence, the amplitude) for
cosmological perturbations of different nature which can be generated
quantum-mechanically in one and the same cosmological model.
A consistent quantum theory provides us, of course, with both,
the mean value of $N$ and its variance.  According to the calculations
of Ref.~[3], the contribution of quantum-mechanically generated
gravitational waves to the large-angular-scale anisotropy is somewhat
greater (even in the limit of the de Sitter expansion) than the
contribution of quantum-mechanically generated density perturbations.
It is argued in~[7,8] that the ``standard'' inflationary formula for
density perturbations, which requires in this limit an arbitrarily
large excess of density perturbations over gravitational waves,
is based on errors.  We additionally discuss this issue in some detail
in the Appendix of this paper. However, the major emphasis of this paper
is on the statistical properties of cosmological perturbations which are
determined by their common origin --- quantum mechanics and squeezing.
These properies are related to the variance of $N$ rather than to its
mean value. Our discussion will be equally well applicable to the
perturbations of any nature (density perturbations, rotational
perturbations, gravitational waves) if they have the same origin.

\section{The General Equations for Quantized Cosmological Perturbations}

Here we will briefly summarize some basic information about quantized
cosmological perturbations (see~[3,7] and references therein).
The squeezed field operator derived in this Section is a basic
mathematical construction for our further discussion of statistical
properties.

The metric of the homogeneous isotropic universe can be written in
the form
\begin{equation}
   ds^2 =-a^2(\eta )(d\eta^2 -\gamma_{ij} \, dx^i \, dx^j ) \quad ,
\end{equation}
where $\gamma_{ij}$ is the spatial metric. For reasons of simplicity,
we will be considering only spatially flat universes, that is
$\gamma_{ij} = \delta_{ij}$.

Following Lifshitz, it is convenient to write the perturbed metric
in the form
\begin{equation}
    ds^2 = -a^2(\eta )
    [d\eta^2 - (\delta_{ij} + h_{ij}) dx^i\, dx^j] \quad ,
\end{equation}
where $h_{ij}$ are functions of $\eta$-time and spatial coordinates.
By writing the perturbed metric in this form we do not lose anything
in the physical content of the problem, but we gain considerably
in the mathematical tractability of the perturbed Einstein equations.
The one who is interested in solving equations
will certainly be interested in their simpler form. Those who prefer
``gauge-invariant formalisms'' are welcome to take the found solution
and compute with its help whichever gauge-invariant quantity they like.
These quantities, being gauge-invariant, have the same values in
all gauges.

The components $h_{ij}$ of the perturbed gravitational field can
be classified in terms of scalar, vector, and tensor eigenfunctions
of the Laplace differential operator. The components of the
perturbed energy-momentum tensor can also be classified in the
same manner. After that, the linearized Einstein equations reduce
to a set of ordinary differential equations, separately for scalar
(density perturbations), vector (rotational perturbations),
and tensor (gravitational waves) parts.

The number of independent unknown functions of time that can
potentially be present (on grounds of the classification scheme)
in the perturbed Einstein equations is always greater than the number
of independent equations. It is 6~functions and 4~equations for density
perturbations, 3~functions and 2~equations for rotational perturbations,
and 2~functions and 1~equation for gravitational waves. In order
to make the system of equations closed, it is necessary to say
something about the perturbed components of the energy-momentum tensor
or to specify from the very beginning the form of the energy-momentum
tensor. The popular choices are perfect fluids and scalar fields.
Even for gravitational waves, it is not a totally trivial question
what their definition is (see, for example, Ref.~[9]). However,
after everything is being set, and as soon as the scale factor
$a(\eta )$ (the background solution) is known, the general solution
to the perturbed equations can be found.  In practice, exact
solutions are being found piecewise, at the intervals of
evolution where the energy-momentum tensor has simple prescribed
forms.

We can now write the quantum-mechanical operator for the
perturbations of the gravitational field $h_{ij}$ in the following
universal form:
\begin{equation}
  h_{ij} = {C\over a(\eta )} {1\over (2\pi)^{3/2}}
  \int_{-\infty}^\infty d^3n \sum_{s=1}^2
  {\stackrel{s}{p}}_{ij} ({\bf n})
  {1\over\sqrt{2n}}
\left[ {\stackrel{s}{c}}_{\bf n}(\eta ) e^{i{\bf nx}}
     + {\stackrel{s}{c}}_{\bf n}^\dagger (\eta )
       e^{-i{\bf nx}} \right] \, .
\end{equation}

We will start the explanation of Eq.~(3) from the polarization
tensors ${\stackrel{s}{p}}_{ij}$.  Let us introduce, in addition
to the unit wave-vector ${\bf n}/n$, two more unit vectors
$l_i$, $m_i$, orthogonal to each other and to ${\bf n}$:
\begin{eqnarray}
   {n_i \over n} ~
&&= (\sin\theta\cos\phi ,\, \sin\theta\sin\phi ,\, \cos\theta )\, ,
    \qquad
   l_i = (\sin\phi ,\, -\cos\phi ,\, 0) \, ,\nonumber\\
   m_i~
&&= \pm (\cos\theta\cos\phi ,\, \cos\theta\sin\phi ,\, -\sin\theta )\, ,
\end{eqnarray}
$+$ for $\theta < {\pi\over 2},~~$
$-$ for $\theta > {\pi\over 2}$.

The two independent polarization tensors, $s=1,2$, for each class of
perturbations, can be written as follows. For gravitational waves:
\[
  {\stackrel{1}{p}}_{ij}({\bf n})=(l_il_j - m_im_j)\, , \qquad
  {\stackrel{2}{p}}_{ij}({\bf n})=(l_im_j + l_jm_i)\, .
\]
For rotational perturbations:
\[
   {\stackrel{1}{p}}_{ij}({\bf n})
 = {1\over n} (l_i n_j + l_j n_i)\, , \qquad
   {\stackrel{2}{p}}_{ij}({\bf n})
 = {1\over n}(m_i n_j + m_j n_i)\, .
\]
For density perturbations:
\[
  {\stackrel{1}{p}}_{ij}({\bf n})
  = \sqrt{{2\over 3}}\, \delta_{ij} \, , \qquad
  {\stackrel{2}{p}}_{ij}({\bf n})
  = -\sqrt{3}\, {n_in_j\over n^2}+{1\over\sqrt{3}}\delta_{ij} \, .
\]
The polarization tensors of each class satisfy the conditions
${\stackrel{s}{p}}_{ij}\, {\stackrel{s'}{p^{ij}}}=2\delta_{ss'}$,
${\stackrel{s}{p}}_{ij}(-{\bf n})={\stackrel{s}{p}}_{ij}({\bf n})$.
In practical handling of the density perturbations it proves
convenient to use sometimes, in addition to the {\it scalar}
polarization component ${\stackrel{1}{p}}_{ij}$, the
{\it longitudinal-longitudinal} component (proportional to
$n_in_j$) instead of ${\stackrel{2}{p}}_{ij}$.
The explicit functional dependence of the polarization tensors
is needed for the calculation of various angular correlation functions.

The evolution of the creation and annihilation operators
${\stackrel{s}{c}}_{\bf n} (\eta )$,
${\stackrel{s}{c}}_{\bf n}^\dagger (\eta )$,
for each class of perturbations and for each polarization state,
is defined by the Heisenberg equations of motion:
\begin{equation}
  {dc_{\bf n}(\eta )\over d\eta}
= -i [c_{\bf n}(\eta ),H] \, , \qquad
  {dc_{\bf n}^\dagger (\eta )\over d\eta}
= -i [c_{\bf n}^\dagger (\eta ),H] \, .
\end{equation}
The dynamical content of the problem is determined by the Hamiltonian
$H$. Its form depends on the class of perturbations and additional
assumptions about the energy-momentum tensor which we have to make,
as was discussed above.

Under the simplest assumptions about gravitational waves (waves
interact only with the background gravitational field, there is
no anisotropic material sources) the Hamiltonian for each
polarization component takes on the form
\begin{equation}
  H = nc_{\bf n}^\dagger c_{\bf n}
    + nc_{-{\bf n}}^\dagger c_{-{\bf n}}
    + 2\sigma (\eta ) c_{\bf n}^\dagger c_{-{\bf n}}^\dagger
    + 2\sigma^\ast (\eta ) c_{\bf n} c_{-{\bf n}}
\end{equation}
where the coupling function $\sigma (\eta )$ is
$\sigma (\eta ) ={i\over 2}\,{a' \over a}$.

For rotational perturbations, assuming that the primeval matter
is capable of supporting torque oscillations, assuming that
the oscillations are minimally coupled to gravity,
and assuming that the torsional velocity of sound is equal
to the velocity of light, the Hamiltonian for each polarization
component reduces to exactly the same form (6) with the same
coupling function $\sigma (\eta )$.

For density perturbations, we consider specifically a minimally
coupled scalar field with arbitrary scalar field potential
as a model for matter in the very early Universe,
and perfect fluids at the later eras. The quantization
is based on the {\it scalar} polarization component
(the function of time responsible for another polarization
state is not independent). There is only one independent
sort of creation and annihilation operators in this case.
The operators
${\stackrel{s}{c}}_{\bf n}(\eta )$,
${{\stackrel{s}{c}}}_{\bf n}^\dagger (\eta )$
are expressible in terms of the operators
$d_{\bf n} (\eta )$, $d_{\bf n}^\dagger (\eta )$
for which the Hamiltonian has again the same form (6)
but with the coupling function
$\sigma (\eta )={i\over 2}{(a\sqrt\gamma )' \over a\sqrt\gamma}$,
where
\[
  \gamma (\eta ) = 1+ \left( {a \over a'} \right)^\prime \quad .
\]
For density perturbations, it is the operators
$d_{\bf n}(\eta )$, $d_{\bf n}^\dagger (\eta )$
that participate in Eqs.~(5), (6).

Now, let us turn to the constant $C$ in Eq.~(3). Its value is
determined by the normalization of the field of each class
to the ``half of the quantum in each mode''.  Under the assumptions
listed above, one derives
$C = \sqrt{16\pi}\, l_{Pl}$ for gravitational waves,
$C = \sqrt{32\pi}\, l_{Pl}$ for rotational perturbations,
and $C = \sqrt{24\pi}\, l_{Pl}$ for density perturbations,
where $l_{Pl}$ is the Planck length,
$l_{Pl} = (G\hbar /c^3)^{1/2}$.

The form of the Hamiltonian (6) dictates the form of the solution
(Bogoliubov transformation) to Eq.~(5):
\begin{eqnarray}
    c_{\bf n} (\eta )~
&&= u_n(\eta ) c_{\bf n}(0)
    + v_n(\eta )c_{-{\bf n}}^\dagger (0)\nonumber\\
    c_{\bf n}^\dagger (\eta )~
&&= u_n^\ast (\eta ) c_{\bf n}^\dagger (0)
  + v_n^\ast (\eta ) c_{-{\bf n}}(0)
\end{eqnarray}
where $c_{\bf n}(0)$, $c_{\bf n}^\dagger (0)$
are the initial values of the operators taken long before
the interaction with the pump field became important
($\sigma (\eta )/n \rightarrow 0$) and which define
the vacuum state $c_{\bf n}(0)|0\rangle =0$.
The complex functions $u_n(\eta )$, $v_n(\eta )$
obey coupled first-order differential equations
following from Eq.~(4) and satisfy the condition
$|u_n|^2 - |v_n|^2 = 1$ which guarantees that
the commutator relationship
$[c_{\bf n}(0), c_{\bf m}^\dagger (0)]=\delta^3({\bf n}-{\bf m})$
is satisfied at all times,
$[c_{\bf n}(\eta ),c_{\bf m}^\dagger (\eta )]=\delta^3({\bf n}-{\bf m})$.
If one introduces the function
$\mu_n(\eta )= u_n(\eta )+v_n^\ast(\eta )$,
one recovers from the equations for
$u_n(\eta )$, $v_n(\eta )$
the classical equations of motion.  For gravitational waves:
\begin{equation}
  \mu_n^{\prime\prime}
+ \left[ n^2 -{a^{\prime\prime}\over a}\right]\mu_n =0 \quad .
\end{equation}
For rotational perturbations:
\begin{equation}
  \mu_n^{\prime\prime}
+ \left[ n^2 {v_t^2 \over c^2}
       -{a^{\prime\prime}\over a}\right]\mu_n =0 \quad .
\end{equation}
where $v_t$ is the torsional velocity of sound which we assumed
above to be $c$. For the scalar field density perturbations:
\begin{equation}
  \mu_n^{\prime\prime}
+ \left[ n^2 - {(a\sqrt\gamma )^{\prime\prime} \over a\sqrt\gamma}
  \right] \mu_n =0 \quad .
\end{equation}
If the pump field is such that the $\gamma$ function is
independent of time, Eq.~(10) reduces to exactly the same form as
Eq.~(8) for gravitational waves.

In the Schr\"odinger picture, the initial vacuum quantum state
$|0_{\bf n}\rangle$ $|0_{-{\bf n}}\rangle$
evolves into a two-mode squeezed vacuum quantum state.
In our problem, each of the two-mode squeezed vacuum quantum
states is a product of two identical one-mode squeezed vacuum
quantum states which correspond to the decomposition of
the real field $h_{ij}$ over real spatial harmonics
$\sin {\bf nx}$ and $\cos {\bf nx}$.
In the Heisenberg picture, the initial vacuum quantum
state does not evolve in time and is the same now.

By using Eq.~(7) one can present the field (3) in the form
\begin{equation}
  h_{ij} (\eta ,{\bf x}) = C{1\over (2\pi)^{3/2}}
\int_{-\infty}^\infty d^3n \sum_{s=1}^2
{\stackrel{s}{p}}_{ij}({\bf n}) {1\over \sqrt{2n}}
\left[ {\stackrel{s}{h}}_n (\eta ) e^{i{\bf nx}}\,
       {\stackrel{s}{c}}_{\bf n} (0)
     + {\stackrel{s}{h}}_n^\ast (\eta ) e^{-i{\bf nx}}\,
       {\stackrel{s}{c}}_{\bf n}^\dagger (0) \right] \, ,
\end{equation}
where the functions
${\stackrel{s}{h}}_n(\eta )$ are
${\stackrel{s}{h}}_n(\eta )= {1\over a(\eta )}
[{\stackrel{s}{u}}_n(\eta )+{\stackrel{s}{v}}_n^\ast (\eta )]$.
For gravitational waves and rotational perturbations, the functions
${\stackrel{s}{h}}_n$ are simply
${\stackrel{s}{h}}_n = {\stackrel{s}{\mu}}_n/a$ where
${\stackrel{s}{\mu}}_n$ are solutions to Eqs.~(8), (9) with
appropriate initial conditions. For density perturbations,
the functions ${\stackrel{s}{h}}_n$ are derivable from solutions
to Eq.~(10) in accord with the relationship between $c$ and $d$
operators. Besides, for density perturbations, we should regard
${\stackrel{1}{c}}_{\bf n} (0)= {\stackrel{2}{c}}_{\bf n}(0)$,
${\stackrel{1}{c}}_{\bf n}^\dagger(0)
={\stackrel{2}{c}}_{\bf n}^\dagger(0)$
in Eq.~(11).  In all cases, for a given cosmological model,
that is for a model in which the scale factor $a(\eta )$
is known from the very early times and up to now,
the functions ${\stackrel{s}{h}}_n$ can be found from the
classical equations of motion with appropriate initial conditions.

It follows from Eq. (11) that the mean quantum-mechanical value
of the field $h_{ij}$ is zero at every spatial point and at
every moment of time, $\langle 0|h_{ij}|0\rangle = 0$.
One can also calculate variances of the field,
that is the expectation values of its quadratic combinations.
One useful quantity is $h_{ij}h^{ij}$.  By manipulating with
the product of two expressions (11), using the summation properties
of the polarization tensors, and remembering that the only
nonvanishing correlation function is
\[
  \langle 0|{\stackrel{s}{c}}_{\bf n} (0)
  {\stackrel{s'}{c}}_{{\bf n}'}^\dagger (0) |0\rangle
= \delta_{ss'} \delta^3({\bf n}-{\bf n}') \quad ,
\]
one can derive the formula
\begin{equation}
\langle 0|h_{ij}(\eta ,{\bf x}) h^{ij}(\eta ,{\bf x})|0\rangle
= {C^2\over 2\pi^2} \int_0^\infty n \sum_{s=1}^2
  |{\stackrel{s}{h}}_n(\eta )|^2 \, dn \quad .
\end{equation}
Equation (12) shows that the variance is independent of the
spatial point ${\bf x}$ but does depend on time.

The expression under the integral in formulas such as Eq.~(12)
is usually called the power spectrum (in this case, it is the
power spectrum of the quantity $h_{ij}h^{ij}$):
\begin{equation}
   P(n) = {C^2\over 2\pi^2} n \sum_{s=1}^2
   |{\stackrel{s}{h}}_n(\eta )|^2 \quad .
\end{equation}
In cosmology, it is common to use the power spectrum defined
in terms of the logarithmic frequency interval, that is
the function
\begin{equation}
   P_Z(n) = {C^2\over 2\pi^2} n^2 \sum_{s=1}^2
   |{\stackrel{s}{h}}_n(\eta )|^2
\end{equation}
($Z$ from Zeldovich). We are mostly interested in the power
spectrum of cosmological perturbations in the present Universe,
at the matter-dominated stage. This spectrum is never smooth
as a function of frequency (wave-number) $n$. Squeezing and
associated standing wave pattern of the field make the spectrum
an oscillating function of $n$ for each moment of time.
In their turn, the oscillations in the power spectrum will
produce oscillations in the distribution of the higher-order
multipoles of the angular correlation function for the temperature
anisotropies.
However, the spectrum is smooth for sufficiently long waves.
At a given moment of time, it applies to all perturbations
whose wavelengths are of the order and longer than the Hubble
radius defined at that time.  Moreover, the smooth part of
the spectrum is power-law dependent on $n$ if the scale factor
$a(\eta )$ of the very early Universe (the pump field) was
power-law dependent on $\eta$-time.

Let us assume that the scale factor at the initial stage of
expansion was
\begin{equation}
  a(\eta ) = l_0 |\eta |^{1+\beta}
\end{equation}
where $l_o$ and $\beta$ are constants. If the evolution
is governed by a scalar field, the Einstein equations require
the constant $\beta$ to be $\beta \leq -2$.  The value
$\beta = -2$ corresponds to the de Sitter expansion.
At later times, the scale factor changed to the laws of the
radiation-dominated and matter-dominated universes.
{}From solutions for ${\stackrel{s}{h}}(\eta )$ traced up
to the matter-dominated stage, one can find
\[
   \sum_{s=1}^2 |{\stackrel{s}{h}}_n(\eta )|^2
   \sim {1\over l_o^2} n^{2\beta +2}
\quad {\rm and}\qquad
   P_Z(n)\sim {l_{Pl}^2 \over l_o^2} n^{2(\beta +2)} \, .
\]
It is convenient to introduce the {\it characteristic} amplitude
$h(n)$ of the metric perturbations defining this amplitude
as the standard deviation (square root of variance) of the
perturbed gravitational field per logarithmic frequency interval.
In the long-wavelength limit under discussion, this quantity
is universally expressed (both, for gravitational waves
and density perturbations) by the formula [21]:
\begin{equation}
  h(n) \sim {l_{Pl} \over l_o} n^{\beta+2} \quad .
\end{equation}
Note that the functional form of $h(n)$ is the same for
gravitational waves and density perturbations,
the difference is in the numerical coefficient
(omitted in this discussion) which is somewhat in favor
of gravitational waves~[3,7]. The numerical level of $h(n)$
is mainly controlled by the constant $l_o$.

The spectra of other quantities can be found in the same manner.
For instance, in case of density perturbations, one can derive
the spectrum of perturbations in the matter density
$\delta\rho /\rho$. Since the relationship between
$\delta\rho /\rho$ and the metric perturbations involves
the factor $(n\eta )^2$ and, hence, involves two extra
powers of $n$, ${\delta\rho \over \rho}(n) \sim n^2h(n)$,
one finds
\[
  \langle 0|
  {\delta\rho \over \rho}\, {\delta\rho \over \rho}
  |0\rangle \sim \int_0^\infty P_Z^\rho(n) {dn \over n}
\]
where
$P_Z^\rho(n) \sim (l_{Pl}^2 /l_o^2)n^{2(\beta +4)}$
and
\begin{equation}
     {\delta\rho \over \rho}(n)
\sim {l_{Pl}\over l_o} n^{\beta +4} \quad .
\end{equation}

It follows from Eq.~(16) that $h(n)$ is independent of $n$ if
$\beta =-2$.  This independence corresponds to the original
Zeldovich's definition of the ``flat'' spectrum:  all waves enter
the Hubble radius with the same amplitude. If the gravitational
field perturbations $h(n)$, regardless of their wavelength,
have equal amplitudes upon entering the Hubble radius,
the matter density perturbations ${\delta\rho\over\rho}(n)$
do also have equal amplitudes (the extra factor $(n\eta )^2$
is of the order of 1 at the time when a given wave $n$ enters
the Hubble radius). For models of the very early Universe
governed by a scalar field, the spectral index $\beta +2$
in Eq.~(16) can never be positive.

Formula (16) and the associated formula (17) should be compared
with the ``standard'' inflationary formula which requires that the
amplitudes of density perturbations taken at the time of entering
the Hubble radius should go to infinity in the limit of the
de Sitter inflation, $\beta\rightarrow -2$. It should also be
noted that a ``disgusting convention'' (the term is borrowed from
Ref.~[10]) is often being used according to which one and the same
Harrison-Zeldovich spectrum is described by the spectral index
$n_t=0$ for gravitational waves and by the spectral index
$n_s=1$ for density perturbations. Of course, there is no need
in this convention. In both cases, the metric perturbations
with the Harrison-Zeldovich spectrum are described by the same
spectral index (zero), see Eq.~(16).

We will finish this section with a short discussion of coherent
states. There is no natural mechanism for the generation
of cosmological perturbations in coherent states,
but if there were one it would be reflected in many parts
of the theory. The interaction part of the Hamiltonian (6)
would be linear (not quadratic) in the creation and annihilation
operators.  The analogue of Eq.~(7) would read
\begin{eqnarray}
    c_{\bf n}(\eta )
&&= e^{-in\eta} c_{\bf n}(0) +\alpha_n(\eta ) \nonumber\\
    c_{\bf n}^\dagger (\eta )
&&= e^{in\eta} c_{\bf n}^\dagger (0) +\alpha_n^\ast (\eta )\quad ,
\end{eqnarray}
where the complex function $\alpha_n(\eta )$ is determined
by the coupling function in the Hamiltonian.
On the position --- momentum diagram, the evolution (18)
of the field operators corresponds to {\it displacing} the vacuum
state without {\it squeezing} whereas the evolution (7) corresponds
to {\it squeezing} the vacuum state without {\it displacing}.
In terminology of mechanics, coherent states are produced
by a force acting on the oscillator whereas squeezed vacuum
states are produced by a parametric influence.
In coherent states, the mean value of the field is not zero.
The correlation functions (at least, for some quantities) would also
be different from the squeezed state case. In cosmological
context, this would
eventually be reflected in the differing statistical properties
of perturbations and induced microwave background anisotropies.

The calculations of the next Section are based in an essential
way on the field operator (11) for the squeezed vacuum perturbations.

\section{Quantum-Mechanical Expectation Values for the Microwave
Background Anisotropies}

The microwave background anisotropies are a subject of intense
study~[22].

In absence of cosmological perturbations, the temperature of the
microwave background radiation seen in all directions on the sky
would be the same, $T$. Let us denote a direction on the sky
by a unit vector ${\bf e}$. The presence of cosmological
perturbations makes the temperature seen in the direction
${\bf e}$ differing from $T$. The temperature perturbation
produced by density perturbations or gravitational waves
can be described by the formula [11]:
\begin{equation}
 {\delta T\over T} ({\bf e}) = {1\over 2} \int_0^{w_1}
{\partial h_{ij} \over \partial\eta} e^ie^j\, dw
\end{equation}
where $\partial h_{ij}/\partial\eta$ is taken along the
integration path $x^i=e^iw$, $\eta = \eta_R-w$,
from the event of reception $w=0$ to the event of emission
$w=w_1=\eta_R-\eta_E$.  The formula for rotational perturbations
is more complicated than (19) [11], and we will leave rotational
perturbations aside.

For quantized cosmological perturbations,
the ${\delta T \over T}({\bf e})$ becomes a quantum-mechanical operator.
Using Eq.~(11) we can write this operator as
\begin{eqnarray}
   {\delta T\over T}({\bf e}) =
&&~{C\over 2} {1\over (2\pi)^{3/2}}
   \int_0^{w_1} dw \int_{-\infty}^\infty d^3n
   \sum_{s=1}^2 {\stackrel{s}{p}}_{ij}({\bf n}) e^ie^j \nonumber\\
&&\times ~\left[ {\stackrel{s}{c}}_{\bf n}(0)
     {\stackrel{s}{f}}_n(w)e^{iw{\bf ne}}
   + {\stackrel{s}{c}}_{\bf n}^\dagger(0)
     {\stackrel{s}{f}}_n^\ast (w) e^{-iw{\bf ne}} \right]
\end{eqnarray}
where
\[
 {\stackrel{s}{f}}_n(w) \equiv {1\over \sqrt{2n}}
 {d{\stackrel{s}{h}}_n\over d\eta}\Bigg|_{\eta=\eta_R-w}\quad .
\]

Having defined the observable ${\delta T\over T}({\bf e})$
and knowing the quantum state $|0\rangle$ we can compute
various quantum-mechanical expectation values.
In the laboratory quantum mechanics, the verification
of theoretical predictions expressed in terms of the
expectation values would require experiments on many identical
systems. An immediate generalization of this principle to
cosmology would require speculations about outcomes of
experiments performed in ``many identical universes''.
Without having access to ``many universes'' we can only
rely on the mean (expected) values of the observables
and on the probability distribution functions as indicators
of what is likely or not to be observed in our own single
Universe. We will return to this point in Sec.~IV.

The expected value of the temperature perturbation to be observed
in every fixed direction on the sky is zero:
\[
   \langle 0| {\delta T\over T} ({\bf e}) |0\rangle = 0 \quad .
\]
One particular measured temperature map is the result of the
measurement performed over one particular realization of
the random process describing cosmological perturbations
of quantum-mechanical origin. For this realization,
the temperature perturbations may, should, and in fact are,
present. Many measurements will not help (except of reducing
the instrumental noises) in the sense that they all should
give identical results, because the timescale of the
perturbations under discussion is so enormously larger
than an interval of time between the experiments.
If the COBE's map is correct, we will have to live with
this map practically forever.

Let us now compute the expected angular correlation function
for the temperature perturbations seen in two given directions
on the sky, ${\bf e}_1$ and  ${\bf e}_2$.  This correlation
function is defined as the mean value for the product of
${\delta T\over T} ({\bf e}_1)$ and
${\delta T\over T} ({\bf e}_2)$:
\begin{equation}
   K ({\bf e}_1, {\bf e}_2)
= \langle 0|
  {\delta T\over T} ({\bf e}_1)
  {\delta T\over T} ({\bf e}_2)|0\rangle \quad .
\end{equation}
By manipulating with the product of two expressions (20)
one can derive the formula
\begin{eqnarray}
   K({\bf e}_1, {\bf e}_2) =
&& {1\over 4} C^2 {1\over (2\pi)^3}
   \int_0^{w_1} dw \int_0^{w_1} d\bar w
   \int_{-\infty}^\infty d^3n \,
   e^{i{\bf n}({\bf e}_1w-{\bf e}_2\bar w)}\nonumber\\
&& \times \sum_{s=1}^2
   \left( {\stackrel{s}{p}}_{ij}({\bf n}) e_1^i e_1^j \right)
   \left( {\stackrel{s}{p}}_{ij}({\bf n})e_2^ie_2^j\right)
   {\stackrel{s}{f}}_n(w)
   {\stackrel{s}{f}}_n^\ast (\bar w )\quad .
\end{eqnarray}
The next step is the formidable task of taking the integrals over
angular variables in 3-dimensional wave-vector ${\bf n}$ space.
However, it can be done (see Ref.~[12] for gravitational waves,
Ref.~[13] for rotational perturbations, and Ref.~[3] for density
perturbations). The final expression reduces, without making
any additional assumptions whatsoever, to the form
\begin{equation}
   K({\bf e}_1,{\bf e}_2) = K(\delta )
 = l_{Pl}^2 \sum_{l=l_{\min}}^\infty
   K_l \, P_l(\cos\delta ) \quad .
\end{equation}
We see that the correlation function depends only on the angle
$\delta$ between the directions ${\bf e}_1, {\bf e}_2$ not
directions themselves. The coefficient $l_{Pl}^2$ is taken from
$C^2$, other numerical coefficients are included in $K_l$.
The quantities $K_l$ involve the integration of
${\stackrel{s}{f}}_n(w)$ over the parameter $w$
and the remaining integration over the wave-numbers $n$.
The numerical values of $K_l$ depend on a chosen sort
of cosmological perturbations and a chosen cosmological model;
so far, the formula (23) is totally general.
$P_l(\cos\delta )$ are the Legendre polynomials.
The lowest multipole $l_{\min}$ follows automatically
from the theory and it turns out to be, not surprisingly,
$l_{\min} = 0$ for density perturbations, $l_{\min} = 2$
for gravitational waves (and $l_{\min} = 1$ for rotational
perturbations). For the separation angle $\delta =0$,
Eq.~(23) reduces to the variance of
${\delta T\over T}({\bf e})$, that is
\begin{equation}
   \langle 0 | {\delta T\over T} ({\bf e})
    {\delta T\over T} ({\bf e}) | 0\rangle
=  K(0) = l_{Pl}^2 \sum_{l=l_{\min}}^\infty K_l \quad .
\end{equation}

Formula (23) gives the expected value of the observable
${\delta T\over T}({\bf e}_1){\delta T\over T}({\bf e}_2)$.
If the experimenter measured this observable in ``many universes''
and averaged the measured numbers, he/she would get the result (23).
Moreover, formula (23) says that if the experimenter made the
measurements at any other pair of directions, but with the same
separation angle $\delta$, he/she would again get, after the
averaging over ``many universes'', the result (23).
Without having access to ``many universes'', we can ask what is
the theoretical standard deviation of the quantity
${\delta T\over T}({\bf e}_1){\delta T\over T}({\bf e}_2)$.
(In practice, for deriving $K(\delta )$ we need a kind
of ergodic hypothesis allowing us to replace the averaging over
``universes'' by the averaging over pixels
on a single map.) The variance $V({\bf e}_1,{\bf e}_2)$
of this quantity is, by definition,
\begin{equation}
  V({\bf e}_1,{\bf e}_2)
= \langle 0|
  {\delta T\over T} ({\bf e}_1)
  {\delta T\over T} ({\bf e}_2)
  {\delta T\over T} ({\bf e}_1)
  {\delta T\over T} ({\bf e}_2) |0\rangle
- \left[ \langle 0|
  {\delta T\over T} ({\bf e}_1)
  {\delta T\over T} ({\bf e}_2)
  |0\rangle \right]^2 \, .
\end{equation}
The standard deviation is the square root of this number.

The calculation of $V({\bf e}_1,{\bf e}_2)$
requires us to deal with the product of four expressions (20).
However, the mean values of the products of four creation
and annihilation operators are easy to handle. One can show
that $V({\bf e}_1,{\bf e}_2)$ depends only on the separation
angle $\delta$ and
\begin{equation}
  V({\bf e}_1,{\bf e}_2) = V(\delta )
= \left[ \langle 0|
  {\delta T\over T} ({\bf e}_1)
  {\delta T\over T} ({\bf e}_2) |0\rangle \right]^2
+ \left[ \langle 0|
  {\delta T\over T} ({\bf e})
  {\delta T\over T} ({\bf e}) |0\rangle \right]^2 \, ,
\end{equation}
that is
\begin{equation}
  V(\delta ) = K^2 (\delta )+ K^2(0) \quad .
\end{equation}
The standard deviation for the observable
${\delta T\over T}({\bf e}_1){\delta T\over T}({\bf e}_2)$ is
\begin{equation}
 \sigma (\delta ) = [V(\delta )]^{1/2}
= \sqrt{K^2(\delta )+ K^2 (0)} \quad .
\end{equation}
In a similar fashion one can derive the higher order correlation
functions for two directions ${\bf e}_1$, ${\bf e}_2$
and the correlation functions for larger number of directions,
but we will not need this information.

In the limit $\delta =0$ Eqs.~(25), (26) say that
\begin{equation}
  \langle 0| \left[ {\delta T\over T}({\bf e})\right]^4 |0\rangle
= 3\left[ \langle 0| \left[ {\delta T\over T}({\bf e})\right]^2
  |0\rangle\right]^2 \quad .
\end{equation}
The familiar factor 3 relating the fourth-order moment with
the square of the second-order moment (given that the first-order
moment is equal to zero) is the reflection of the underlying
Gaussian nature of the squeezed vacuum wavefunctions associated
with the Hamiltonian (6).

By examining Eq.~(28) one can conclude that for each separation
angle $\delta$ the standard deviation of the angular correlation
function is very big.  Even at those separation angles at which
$K(\delta )$ vanishes, the standard deviation is as big
as the variance for ${\delta T\over T}({\bf e})$ itself. However,
the value of the standard deviation for a given variable is
not very informative {\it per se}, as long as the probability density
function for this variable is not known. If the probability
density function (p.d.f.) were normal, we could say that
the probability to find a result outside of $1\sigma$ interval
is 32\%. Without knowing the p.d.f. we could resort to the
Chebyshev inequality, but it would only tell us that this
probability is less than 1.  To get more information about
possible deviation of the angular correlation function
from its mean values we will consider in Sec.~IV a classical
random model which will reproduce the expectation values
calculated above and will allow us to construct the p.d.f.
for the variable
${\delta T \over T}({\bf e}_1)$ ${\delta T\over T}({\bf e}_2)$.
On the other hand, the quantum-mechanical calculations
of this Section will shed light on the classical model.
As is known, ``quantum mechanics helps us understand classical
mechanics'', see on this subject a paper of Zeldovich signed
by the pseudonym Paradoksov~[14].

\section{Classical model for the statistics of the microwave
background anisotropies}

A distribution of the microwave background temperature over the
sky is a real function of the angular coordinates. Assuming that
$\delta T/T$ is a sufficiently smooth function on a sphere,
one can expand it over the set of orthonormal complex spherical
harmonics $Y_{lm}(\theta ,\phi )$~[15]:
\begin{equation}
 {\delta T \over T}({\bf e}) = \sum_{l=0}^\infty \sum_{m=-l}^l
[a_{lm}Y_{lm}({\bf e}) + a_{lm}^\ast Y_{lm}^\ast ({\bf e})] \quad .
\end{equation}
We want to formulate a statistical hypothesis about the coefficients
$a_{lm}$, so it is better to write them first in terms of real $(r)$
and imaginary $(i)$ components:
\begin{eqnarray}
    a_{lm}~
&&= a_{lm}^r +i a_{lm}^i \, , \qquad
    a_{lm}^\ast = a_{lm}^r - i a_{lm}^i \, \nonumber\\
    Y_{lm}~
&&= Y_{lm}^r + iY_{lm}^i \, , \qquad
    Y_{lm}^\ast = Y_{lm}^r - iY_{lm}^i \, , \nonumber\\
{\delta T\over T}({\bf e})~
&&= 2\sum_{l=0}^\infty \sum_{m=-l}^l
    [a_{lm}^r Y_{lm}^r ({\bf e})
    - a_{lm}^i Y_{lm}^i ({\bf e})] \quad .
\end{eqnarray}

Our statistical hypothesis is as follows: (i)~all members of the set
of random variables $\{ a_{lm}^r,a_{lm}^i\}$ are statistically
independent, (ii)~each individual variable is normally distributed
and has a zero mean, (iii)~all variables with the same index $l$
have the same standard deviation $\sigma_l$.  All said is expressed
by the probability density function (p.d.f.) for individual variables:
\begin{equation}
  f(a_{lm}^r) = {1\over\sqrt{2\pi}\, \sigma_l}
  e^{-{(a_{lm}^r)^2\over 2\sigma_l^2}} \, , \quad
  f(a_{lm}^i) = {1\over\sqrt{2\pi}\, \sigma_l}
  e^{-{(a_{lm}^i)^2\over 2\sigma_l^2}} \, ,
\end{equation}
and by the p.d.f. for the entire set of variables, which is simply
a product of all p.d.f.'s for all individual variables:
\begin{equation}
   f\left( \{ a_{lm}^r, a_{lm}^i\}\right)
= {\sqcap}_{l,m} f(a_{lm}^r)f(a_{lm}^i) \quad .
\end{equation}

Having postulated the p.d.f.'s, we can now compute the expectation
values of certain functions of the random variables. Below,
the angular brackets will denote the expectation values calculated
with the help of the p.d.f. (33), unless other definition is stated.

Obviously, all linear functions have a zero mean:
\begin{equation}
 <\! a_{lm}^r \!> ~= 0 \, , \qquad <\! a_{lm}^i \!> ~= 0\quad .
\end{equation}
For quadratic combinations we have
\begin{eqnarray}
    <\! a_{l_1m_1}^r a_{l_2m_2}^r\!>~
&&= \sigma_{l_1}^2 \delta_{l_1l_2} \delta_{m_1m_2} \, , \quad
    <\! a_{l_1m_1}^i a_{l_2m_2}^i\!>~
  = \sigma_{l_1}^2 \delta_{l_1l_2} \delta_{m_1m_2} \, , \nonumber\\
    <\! a_{l_1m_1}^r a_{l_2m_2}^i\!> ~
&&= 0 \, , \qquad\qquad\qquad
    <\! a_{l_1m_1}^i a_{l_2m_2}^r\!> ~= 0 \, .
\end{eqnarray}
All triple products have zero means. Among quartic combinations,
only those can survive which have four indices $(r)$, or four
indices $(i)$, or two indices $(r)$ and two indices $(i)$.
Two representative expressions are:
\begin{eqnarray}
    <\! a_{l_1m_1}^r a_{l_2m_2}^r a_{l_3m_3}^r a_{l_4m_4}^r\!>
&&= \sigma_{l_1}^2 \sigma_{l_3}^2
    \delta_{l_1l_2} \delta_{m_1m_2}
    \delta_{l_3l_4} \delta_{m_3m_4} \nonumber\\
  +~\sigma_{l_1}^2 \sigma_{l_2}^2
    \delta_{l_1l_3} \delta_{m_1m_3}
    \delta_{l_2l_4} \delta_{m_2m_4}
&&+ ~\sigma_{l_1}^2 \sigma_{l_2}^2
    \delta_{l_1l_4} \delta_{m_1m_4}
    \delta_{l_2l_3} \delta_{m_2m_3} \quad ,\\
    <\! a_{l_1m_1}^r a_{l_2m_2}^r a_{l_3m_3}^i a_{l_4m_4}^i\!>
&&= \sigma_{l_1}^2 \sigma_{l_3}^2
    \delta_{l_1l_2} \delta_{m_1m_2}
    \delta_{l_3l_4} \delta_{m_3m_4} \quad .
\end{eqnarray}
Other quartic combinations can be obtained by the replacement
$(r)\leftrightarrow (i)$ in Eqs.~(36), (37) (or by permutation
of pairs $(lm)$ in case of Eq.~(37)).  The higher-order correlations
can be derived in a similar way, but we will not need them.

In our further calculations related to the random variables
${\delta T\over T}({\bf e})$ and
${\delta T\over T}({\bf e}_1) {\delta T\over T}({\bf e}_2)$
it is easier to deal with the complex coefficients $a_{lm}$,
so we will first translate the above relationships to them.
By using the available information one can derive
\begin{eqnarray}
    <\! a_{lm} \!> ~
&&= 0 \,, \quad <\! a_{l_1m_1} a_{l_2m_2}^\ast \!> ~
    = 2\sigma_l^2 \delta_{l_1l_2} \delta_{m_1m_2} \quad ,\nonumber\\
  <\! a_{l_1m_1} a_{l_2m_2}
      a_{l_3m_3}^\ast a_{l_4m_4}^\ast \!>~
&&= 4\sigma_{l_1}^2 \sigma_{l_2}^2
    (\delta_{l_1l_3} \delta_{m_1m_3}
     \delta_{l_2l_4} \delta_{m_2m_4}
   + \delta_{l_1l_4} \delta_{m_1m_4}
     \delta_{l_2l_3} \delta_{m_3m_3} ) \, .
\end{eqnarray}
The mean values of the complex conjugated quantities are given
by the same formulas (38). Other nonvanishing quartic combinations
can be obtained from the one in Eq.~(38) by the permutation of
pairs $(lm)$.

Now, even before deriving the p.d.f.'s for the random variables
${\delta T\over T} ({\bf e})$ and
${\delta T\over T} ({\bf e}_1) {\delta T\over T} ({\bf e}_2)$,
we can find some expectation values. It is clear from the
definition (30) and Eq.~(38) that
\begin{equation}
  \Bigg< {\delta T\over T} ({\bf e})\Bigg> = 0   \quad .
\end{equation}
When calculating the angular correlation function one should
remember that
\begin{equation}
  \sum_{m=-l}^l Y_{lm}({\bf e}_1) Y_{lm}^\ast ({\bf e}_2)
= {2l+1\over 4\pi} P_l(\cos\delta )
\end{equation}
(note the origin of the factor $2l+1$ which will accompany us often).
By taking the product of two expressions (30) and using
Eqs.~(38), (40) one can find the angular correlation function
\begin{equation}
  \Bigg< {\delta T\over T} ({\bf e}_1)
    {\delta T\over T} ({\bf e}_2) \Bigg>
= {1\over\pi} \sum_{l=0}^\infty
  \sigma_l^2(2l+1)P_l(\cos\delta ) \quad .
\end{equation}
If the separation angle $\delta$ is zero, we obtain
\begin{equation}
  \Bigg< {\delta T\over T} ({\bf e})
    {\delta T\over T} ({\bf e})\Bigg>
= {1\over\pi} \sum_{l=0}^\infty \sigma_l^2(2l+1) \quad .
\end{equation}
[One may notice an incidental fact that the mean value of the
random variable $a_l^2$ defined as
$a_l^2 =\sum_{m=-l}^l a_{lm}a_{lm}^\ast$ is
$<\! a_l^2 \!> = 2(2l+1)\sigma_l^2$,
that is the same expression which enters Eq.~(41).
This may suggest an interpretation of the quantity
$<\! a_l^4 \!> - <\! a_l^2 \!>^2$
as the variance of the multipole moments. One should be carefull
with this interpretation, see Sec.~V.]

We can also find the 4th order expectation values.
The product of 4 expressions (30) in conjunction with
Eq.~(38) gives
\begin{eqnarray}
&& \Bigg< {\delta T\over T} ({\bf e}_1)
    {\delta T\over T} ({\bf e}_2)
    {\delta T\over T} ({\bf e}_1)
    {\delta T\over T} ({\bf e}_2)\Bigg>
 - \Bigg< {\delta T\over T} ({\bf e}_1)
    {\delta T\over T} ({\bf e}_2)\Bigg>^2\nonumber\\
&&= \Bigg< {\delta T\over T} ({\bf e}_1)
    {\delta T\over T} ({\bf e}_2) \Bigg>^2
  + \Bigg< {\delta T\over T} ({\bf e})
    {\delta T\over T} ({\bf e}) \Bigg>^2 \, .
\end{eqnarray}
If $\delta =0$, it follows from Eq.~(43) that
\begin{equation}
   \Bigg< \left[ {\delta T\over T}({\bf e})\right]^4\Bigg>
= 3\left< \left[ {\delta T\over T}({\bf e})\right]^2\right>^2 \quad .
\end{equation}

Up to difference in the meaning of the angular brackets,
the formulas (39), (43), (44) reproduce the analogous results
of the previous Section. Moreover, from comparison of
Eqs.~(23), (24) with Eqs.~(41), (42) we can relate the
quantities $K_l$, derivable from a given cosmological
model plus perturbations, with the abstract quantities
$\sigma_l$.

We can now engage in our major enterprise --- the construction
of the p.d.f. for the random variable
$v\equiv {\delta T\over T}({\bf e}_1){\delta T\over T}({\bf e}_2)$.
We will start from the p.d.f. for the random variable
$z\equiv {\delta T \over T}({\bf e})$. When it is necessary
to distinguish directions ${\bf e}_1$ and ${\bf e}_2$,
we will use the notations $z_1$ and $z_2$.

The variable $z$ is a function of the variables
$\{ a_{lm}^r, a_{lm}^i\}$  whose p.d.f.'s are known,
Eqs.~(31), (32). There exist regular methods (see, for example,
an excellent book~[16]) allowing to derive rigorously the
p.d.f. of a function. However, in our case that the function
is linear and all p.d.f's are normal, we can partially rely
on a guesswork. Combining formulas and guessing we can write
\begin{equation}
  f(z) = {1\over \sqrt{2\pi} \, \sigma_z}
  e^{-{z^2\over 2\sigma_z^2}} \quad ,
\end{equation}
where
\begin{equation}
  \sigma_z^2 = {1\over \pi} \sum_{l=0}^\infty
  \sigma_l^2 (2l+1) \quad .
\end{equation}
The p.d.f. (45) certainly leads to Eqs.~(39), (42), (44).
Moreover, it allows us to say that the probability to find $z$
outside of $1\sigma_z$ interval is approximately 32\%:
\[
  P(|z| > \sigma_z) \approx 0.32 \quad .
\]

We now introduce two variables, $z_1$ and $z_2$,
and ask about the p.d.f. in the 2-dimensional space
$(z_1,z_2)$.  Again, partially relying on a guesswork,
we find that
\begin{equation}
 f(z_1,z_2) = {1\over 2\pi\sigma_z^2 \sqrt{1-\rho^2}}
\exp \left\{ -{1\over 2\sigma_z^2(1-\rho^2)}
[z_1^2 +z_2^2 -2\rho z_1z_2] \right\}
\end{equation}
where
\begin{equation}
  \rho\sigma_z^2 = {1\over\pi} \sum_{l=0}^\infty
  (2l+1)\sigma_l^2 P_l(\cos\delta ) \, ,\qquad
  | \rho | \leq 1 \quad .
\end{equation}
(See Eq.~(5.11.1) in Ref.~[16]). First, we can check that
the marginal distributions are correct. For $f(z_1)$,
one obtains
\[
  f(z_1) = \int_{-\infty}^\infty f(z_1,z_2)dz_2
= {1\over \sqrt{2\pi}\, \sigma_z}
  e^{-{z_1^2\over 2\sigma_z^2}}
\]
and one obtains a similar expression for $f(z_2)$.
Second, one can check that
\[
  \langle z_1^2 \rangle = \sigma_z^2 \, , \qquad
  \langle z_2^2 \rangle = \sigma_z^2 \, , \qquad
  \langle z_1z_2 \rangle = \rho\sigma_z^2
\]
where the angular brackets mean the integration with the
p.d.f.\ (47). These equalities are Eqs.~(42), (41) which
we must have obtained.

Finally, we shall derive the p.d.f. for the variable $v=z_1z_2$.
We will do this in some detail following the prescriptions of~[16].

Let us introduce the two new variables $(z_1,v)$ instead of
$(z_1,z_2)$ according to the transformation
\[
  z_1 = z_1 \, , \qquad z_2 = {v\over z_1} \quad .
\]
The Jacobian of this transformation is ${\cal J}=1/z_1$.
The p.d.f. $f(v)$ is the result of the following integration:
\[
   f(v) = {1\over 2\pi\sigma_z^2\sqrt{1-\rho^2}}
   \int_{-\infty}^\infty {1\over |z_1|}
   \exp \left\{ -{1\over 2\sigma_z^2(1-\rho^2)}
   \left[ z_1^2+{v^2\over z_1^2}-2\rho v\right]\right\} dz_1 \, .
\]
The integral over $z_1$ can be taken with the help of 3.471.9
from~[17]. The resulting p.d.f. can be written in the form:
\begin{equation}
        f(v) =
\cases{
        {1\over \pi\sigma_z^2\sqrt{1-\rho^2}}
        e^{{\rho v\over \sigma_z^2(1-\rho^2)}}
        K_0\left( {v\over\sigma_z^2(1-\rho^2)}\right) \, ,
&for $v > 0$\cr
        {1\over \pi\sigma_z^2\sqrt{1-\rho^2}}
        e^{{\rho v\over \sigma_z^2(1-\rho^2)}}
        K_0\left( {-v\over\sigma_z^2(1-\rho^2)}\right) \, ,
&for $v < 0$\cr}
\end{equation}
where $K_0$ is the modified Bessel function of its
argument~[18].

The function $f(v)$ is quite complicated and the distribution
is obviously not normal. The function $f(v)$ goes to zero for
$v\rightarrow \pm \infty$ and diverges logarithmically at
the point $v=0$. Even a verification of the normalization
condition
\begin{equation}
    \int_{-\infty}^\infty f(v)dv = 1
\end{equation}
is not trivial. However, with the help of 6.621.3, 9.131.1, 9.121.7,
1.624.9 and 1.623.2 from~[17] one can prove the validity of Eq.~(50).

The mean value and the standard deviation of the variable $v$ are
known, see Eqs.~(41), (43):
\begin{equation}
 \langle v \rangle = \rho\sigma_z^2 \, , \qquad
  \sigma_v
= \left[ \langle v^2\rangle - \langle v\rangle^2\right]^{1/2}
= \sigma_z^2 \sqrt{\rho^2+1} \quad .
\end{equation}
We already knew that the standard deviation is big.
We now see again that $\sigma_v=\sigma_z^2$ at the separation
angles at which the angular correlation function vanishes,
$\rho =0$, and $\sigma_v=\sqrt 2\, \sigma_z^2$ at zero separation
angle, $\rho =1$.  For other separation angles, $\sigma_v$ lies
between these two numbers.

Now that we know the p.d.f., we can assign probabilities to the
different ranges of the variable $v$. For instance, we can
calculate the probability that the measured $v$ will be found,
say, outside of the $\lambda\sigma_v$ interval surrounding
the mean value of $v$, where $\lambda$ is an arbitrary fixed
number.  The probability of our interest is
\begin{equation}
 P(|v-\langle v \rangle | > \lambda\sigma_v)
= \int_{-\infty}^{\sigma_z^2(\rho-\lambda\sqrt{\rho^2+1})}f(v)dv
+ \int_{\sigma_z^2(\rho+\lambda\sqrt{\rho^2+1})}^{\infty}
  f(v)dv\quad .
\end{equation}
To get a qualitative estimate of the associated theoretical
uncertainties for the observable $v$, we will ask a slightly
different question. What should the number $\lambda$ be
in order to have the 0.32 chance of finding $v$ outside
the $\lambda\sigma_v$ interval and, hence, the 0.68 chance
to find it inside the interval?

To evaluate the size of the disaster, we will start from
the case $\rho =0$. In this case, the p.d.f.~(49) is symmetric
with respect to the origin $v=0$ (this is why
$\langle v \rangle$ is zero in this case) and
\begin{equation}
   P(|v| > \lambda\sigma_z^2)
= {2\over \pi} \int_\lambda^\infty K_0(x)dx \quad .
\end{equation}
We want this number to be approximately equal to 0.32.
Judging from the Fig.~9.7 in Ref.~[18], a half of the area
under the $K_0(x)$ function is accumulated when integrating
from approximately $x=1/2$ and up to infinity.
This means that $\lambda$ should approximately be equal
to $1/2$ .

If $\rho \neq 0$ the evaluation of $P$ is more complicated.
For $\rho \neq 0$, the function (49) is not symmetric with
respect to the origin $v=0$.  It has larger values at positive
$v$'s if $\rho > 0$ (this is why $\langle v\rangle > 0$ in
this case) and it has larger values at negative $v$'s if
$\rho < 0$ (this is why $\langle v\rangle < 0$ in this case).
The graph of the function $e^x K_0(x)$ plotted on Fig.~9.8
in Ref.~[18] is helpful. A qualitative analysis shows again
that $\lambda$ is approximately equal to $1/2$. (More accurate
estimates can of course be reached by numerical methods.)

At any rate, the ${1\over 2} \sigma_v$ interval gives
approximately the same probability estimates as if
the distribution (49) were normal.

\section{On the ``cosmic variance''}

The set of random variables $\{ a_{lm}^r,a_{lm}^i\}$
defined by Eqs.~(32), (33) lives its own independent life
regardless of whether or not the variables are considered
random coefficients in the expansion of some function over
spherical harmonics. Being such, it allows introduction
of new functions and calculation of their expectation values.
One interesting variable is defined by the equation
\begin{equation}
  a_l^2 = \sum_{m=-l}^l a_{lm}a_{lm}^\ast
= \sum_{m=-l}^l | a_{lm}|^2
= \sum_{m=-l}^l \Big[ (a_{lm}^r)^2
  + (a_{lm}^i)^2\Big] \, .
\end{equation}
By using Eq.~(38) one can calculate the expectation value of
$a_l^2$:
\begin{equation}
   \langle a_l^2\rangle = (2l+1)2\sigma_l^2 \quad .
\end{equation}
The factor $2(2l+1)$ reflects the number of independent
``degrees of freedom'' associated with the index $l$.
One can also introduce the variable $a_l^4$ and calculate
its expectation value:
\begin{equation}
 \langle a_l^4\rangle = (2l+1)(l+1)8\sigma_l^4
=\langle a_l^2\rangle^2 {2(l+1)\over 2l+1} \quad .
\end{equation}
The difference $\langle a_l^4\rangle -\langle a_l^2\rangle^2$
is, by definition, the variance of the variable $a_l^2$.
{}From Eqs. (56), (55) one finds
\begin{equation}
  \langle a_l^4\rangle - \langle a_l^2\rangle^2
= {1\over 2l+1} \langle a_l^2\rangle^2 \quad .
\end{equation}

This formula, as it stands, expresses a well-known fact: the variance
of the random variable $\chi^2$ defined as the sum of squares of
$n$ independent random variables (degrees of freedom) with the
same normal density, is $n \over 2$ times smaller than the
square of the mean value of $\chi^2$ ~[16]. There is nothing
``cosmological'' or ``inflationary'' in this fact. Knowing the
p.d.f.`s for the set $\{ a_{lm}^r,a_{lm}^i\}$
one can calculate the higher-order correlation functions
for the variable $a_l^2$  and its distribution function~[23-29].
In the recent literature, formula (57) became known as the ``cosmic
variance''. Formula (57) and
a possibility (or lack of) to extract complete information about
a stochastic process from its single realization are, in general,
different issues. For ergodic processes, the existence of a definitely
true relationship (57) prevents in no way the extraction of complete
information about the process from a single realization~[30].

It is important to realize that it is the mean value of the random
variable $a_l^2$, not $a_l^2$ itself, that enters the expected angular
correlation function in front of the Legendre polynomials
and which is often called the multipole moment.
The $a_l^2$ is a random variable and its variance
has meaning, the $\langle a_l^2\rangle$
is a number and its variance has no meaning. Specifically,
one can notice that the angular correlation function (41)
can be written in the form
\begin{equation}
\Bigg< {\delta T\over T}({\bf e}_1)
      {\delta T\over T}({\bf e}_2)\Bigg>
 = {1\over 2\pi} \sum_{l=0}^\infty \langle a_l^2\rangle
    P_l(\cos\delta )\quad .
\end{equation}
On this ground, there may be a temptation
to write the random variable
${\delta T\over T}({\bf e}_1){\delta T\over T}({\bf e}_2)$
in the form
\begin{equation}
  {\delta T\over T}({\bf e}_1){\delta T\over T}({\bf e}_2)
= {1\over 2\pi} \sum_{l=0}^\infty a_l^2 P_l(\cos\delta )
\end{equation}
and to interpret Eq.~(57) as the variance for the multipole
moments of the correlation function. One should
resist to this temptation.

Let us show that the definition (59) is incorrect despite
the fact that it gives correct expectation value (58).
It follows from the definition (59) that
\begin{eqnarray}
&& {\delta T\over T}({\bf e}_1){\delta T\over T}({\bf e}_2)
   {\delta T\over T}({\bf e}_1){\delta T\over T}({\bf e}_2) \nonumber\\
&&= {1\over 4\pi^2} \sum_{l=0}^\infty a_l^4 [P_l(\cos\delta )]^2
  + {1\over 4\pi^2} \sum_{l,l'=0,l\neq l'}^\infty
    a_l^2\, P_l(\cos\delta )a_{l'}^2\, P_{l'}(\cos\delta )\, .
\end{eqnarray}
Using (56), (58) and remembering that $a_l^2$ and $a_{l'}^2$
are statistically independent for $l\neq l'$, one can find
the expectation value of the quantity (60):
\begin{eqnarray}
&& \left< {\delta T\over T}({\bf e}_1){\delta T\over T}({\bf e}_2)
   {\delta T\over T}({\bf e}_1)
   {\delta T\over T}({\bf e}_2) \right>\nonumber\\
&&= {1\over 4\pi^2} \sum_{l=0}^\infty \langle a_l^4\rangle
    [P_l(\cos\delta )]^2
   +{1\over 4\pi^2}
   \sum_{l,l'=0,l\neq l'}^\infty
   \langle a_l^2\rangle \langle a_{l'}^2\rangle
    P_l(\cos\delta )P_{l'}(\cos\delta )\nonumber\\
&&= {1\over 4\pi^2} \sum_{l=0}^\infty
   \langle a_l^4\rangle [P_l(\cos\delta )]^2
   + {1\over 4\pi^2} \left[ \sum_{l=0}^\infty
     \langle a_l^2\rangle P_l(\cos\delta )\right]^2
   - {1\over 4\pi^2} \sum_{l=0}^\infty \langle a_l^2\rangle^2
     [P_l(\cos\delta )]^2 \nonumber\\
&&= \left< {\delta T\over T}({\bf e}_1)
           {\delta T\over T}({\bf e}_2)\right>^2
    +{1\over 4\pi^2}\sum_{l=0}^\infty {1\over 2l+1}
    \langle a_l^2\rangle^2 [P_l(\cos\delta )]^2 \, .
\end{eqnarray}
It follows from (61) that the variance of the variable
${\delta T\over T}({\bf e}_1){\delta T\over T}({\bf e}_2)$
would read (if (59) were correct):
\begin{equation}
  {1\over\pi^2} \sum_{l=0}^\infty (2l+1)\sigma_l^4
  [P_l(\cos\delta )]^2 \quad .
\end{equation}
This expression should be compared with the correct variance following
from Eq.~(43):
\begin{equation}
       {1\over \pi^2}
\left[ \sum_{l=0}^\infty (2l+1)\sigma_l^2 \,
       P_l(\cos\delta)\right]^2
     + {1\over \pi^2}
  \left[ \sum_{l=0}^\infty (2l+1)\sigma_l^2 \right]^2 \, .
\end{equation}
Formulas (62), (63) disagree even for $\delta =0$,
and even in their first, $l=0$ term. This shows that the {\it ad hoc}
definition (59) is incorrect.  The correct definition of the random
variable ${\delta T\over T}({\bf e}_1){\delta T\over T}({\bf e}_2)$
is the one following from the definition (30) and which we have used
in this paper.

\section{Conclusions}

A particular cosmological model plus perturbations gives
unambiguous predictions with regard to the expectation
values of the measurable quantities.  Differing models
give different predictions. We want to distinguish them
observationally and to learn about physics of the very
early Universe. However, the quantum-mechanical origin
of the cosmological perturbations is reflected in the
theoretical statistical uncertainties surrounding the
expectation values. One important measurable quantity
is the angular correlation function of the microwave
background anisotropies. Its mean value at the zero
separation angle was denoted $\sigma_z^2$ in this paper.
It was shown that the standard deviation for the correlation
function is very big. The 68\% confidence level corresponds,
approximately, to ${1\over 2}\sigma_z^2$ at the separation
angles where the correlation function vanishes, to
${\sqrt 2 \over 2}\sigma_z^2$ at the zero separation angle,
and to intermediate numbers for other separation angles.

The angular correlation function has actually been measured.
It is presented at Fig.~3 in the paper~[19].
The authors surround the measured points by a narrow shaded
region which they address as follows: ``The shaded region is
the 68\% confidence region~....~including cosmic variance and
instrument noise''. It is not quite clear what the authors
of Ref.~[19] (see also Ref.~[20]) mean by ``cosmic variance'',
but if they mean the theoretical statistical uncertainties
for the correlation function variable $v$,
these uncertainties are significantly larger than what is plotted.
According to the calculations presented above, the half-width
of the shaded region should be approximately 600~$(\mu K)^2$
near the points where the correlation function vanishes and
approximately 840~$(\mu K)^2$ near the point marking the
zero separation angle.

The conclusion is a bit disappointing. Apparently, God is
telling us something important about the very early Universe
by exhibiting the microwave background anisotropies,
but the {\it channel of information} is so noisy that it will
be hard to understand the message.
\newpage
\section*{appendix. On the ``standard'' inflationary formula for density
perturbations}

In the body of this paper we have studied the statistical properties
of cosmological perturbations of quantum-mechanical origin
and related anisotropies in the microwave background. These properties are
essentially universal, they are equally well true for perturbations
of any nature - density perturbations, rotational perturbations, or
gravitational waves. However, in practice, it is very important to know
which sort of perturbations we are actually dealing with, assuming
that the observed large-angular-scale anisotropy is indeed caused
by perturbations of this
origin. In other words, what does theory say about the comparative
values of the amplitudes, if we agree on the generating mechanism,
equations, and a class
of cosmological models of the very early Universe, say, governed by
the scale factors (15)? In addition to gravitational waves, density
perturbations can also be generated by the same mechanism,
if one makes
favorable assumptions about the dominant matter in the very early Universe
(scalar field) and its coupling to gravity (minimal, the same as for
gravitational waves). According to calculations of Ref.~[3],
density perurbations and gravitational waves will have amplitudes of
the same order of magnitude, wheras according to the ``standard''
inflationary formula the amplitude of density perturbations will be
many orders of magnitude larger than the amplitude of gravitational waves,
if the expansion rate of the very early Universe was sufficiently
close to the archetype inflationary model - the de Sitter expansion.

It is necessarry to say that the quantum-mechanical
generating mechanism has become very popular in the context of the
inflationary hypothesis. Inflationary literature often speaks
about cosmological perturbations being generated ``from
quantum fluctuations''.  However, this literature
associates the explanation of the phenomenon with such things
as ambiguity in the choice of time in the de Sitter universe,
horizon temperature, tremendous inflation of scales, and so on.
The basic concepts are adjusted accordingly. Instead of
{\it amplification}, with the emphasis on a nonvanishing
parametric coupling, increase of amplitude at the expense
of energy of the pump field, quantum-mechanical generation
of waves (particles) in strictly correlated pairs, etc.,
inflationary literature speaks about {\it magnification},
with the emphasis on ``stretching the waves'' and
``crossing the horizons''. Apparently for these reasons,
inflationists did not get puzzled
with their ``standard'' formula for density perturbations
which states that one can produce arbitrarily large amount
of density perturbations by practically doing nothing.

The ``standard'' formula relates the amplitude of density perturbations
today with the values of the scalar field during inflation.
Let us consider, for definiteness, perturbations of the matter
density $\delta\rho /\rho$ with today's wavelengths of the order
of today's Hubble radius $l_H$. The ``standard'' formula says that
\begin{equation}
 {\delta\rho \over \rho}\Bigg|_H \sim {H^2\over \dot\phi (t_i)}
\end{equation}
where the right hand side of this formula is supposed to be evaluated
at the time $t_i$ when the wavelengths of our interest were ``crossing
the horizon'' during inflation. Let us agree with the so-called
``slow-roll'' approximation and assume that the Hubble parameter
$H$ was almost constant during that epoch, $|\dot H| \ll H^2$.
Let us take the numerical value of $H$ during that epoch at the
level, say, 20 orders of magnitude smaller than the Planck value
of $H$. For quantum-mechanically generated gravitational waves,
this would result in today's amplitude $h\approx 10^{-20}$
and the induced anisotropies of the microwave background
$\delta T/T \approx 10^{-20}$ which are much much lower than
the level currently discussed in the experiment. However,
for density perturbations, according to the ``standard''
inflationary formula, the situation is totally different.
Without changing
anything in the curvature of the space-time responsible
for the generating process (that is, leaving $H$ almost
constant and at the same numerical level 20 orders of
magnitude smaller than the Planck value), but simply sending
$\dot\phi (t_i)$ to zero (which corresponds, due to the Einstein
equations, to sending $\dot H$ to zero, {\it i.e.,} making
the ``slow-roll'' approximation better and better,
making the expansion law closer and closer to the
de Sitter expansion and making the shape of the generated
spectrum closer and closer to the scale-invariant form) one
produces arbitrarily large
$(\delta\rho /\rho )|_H$. Inflationists love to stress that
the de Sitter gravitational pump field generates perturbations
with the Harrison-Zeldovich (scale-invariant, flat) spectrum.
What they do not stress is that, according to the ``standard''
formula for density perturbations, the amplitudes of the
scale-invariant spectrum are infinite, and the amplitudes of the
{\it almost} scale-invariant spectrum are {\it almost} infinite.
Instead of blaming their own formula, inflationists blame the
scalar field potentials. This formula is the reason for
rejecting certain scalar field potentials on the grounds
that they generate ``too much'' of density perturbations,
for claims that the contribution of gravitational waves to
$\delta T/T$ is ``negligibly small'' in the limit of the de Sitter
expansion, and even for claims about copious production of black
holes during inflation.

Recently, the ``standard'' formula has been claimed to be confirmed~[31]
and reconfirmed~[32]. In Ref.~[31], this formula
has been formulated, essentially, as the following
``standard result'': ``...~we see that the scalar perturbations can be
very strongly amplified'' [the increase of numerical value from
(almost) zero to (almost) infinity] ``in the course of the transition''
[the instantaneous change of the cosmological scale factor from
one power-law behavior to another power-law behavior]. The authors
of the paper~[31] assure the trusting reader:  ``We think that there
is nothing strange about this~...''.

Here, we will try to understand the origin and mathematical
justification for the ``standard'' inflationary formula.

The early papers which are usually quoted in this connection are
the papers~[33-35]. We will start from the paper
of Hawking~[33] which seems to be clearer than others
in expressing the basic idea and intentions.
The papers~[33-35] are similar in many respects.

Hawking considers a scalar field $\phi$ running slowly down
an effective scalar field potential. He discusses the
inhomogeneous fluctuations
$\phi_1(t,{\bf x})$
in the field
$\phi=\phi_0(t)+\phi_1(t,{\bf x})$
which mean that on a surface of constant time there will be some
regions where the $\phi$ field has run further down the hill
than in other regions. He introduces a new time coordinate
$\bar{t}=t+\delta t(t,{\bf x})$
in such a way that the variations of the field are removed
and the surfaces of constant time are surfaces of constant $\phi$.
Since the scalar field transforms as\\
$\phi_0+\phi_1 \rightarrow \phi_0+\phi_1 - \dot{\phi}_0 \delta t$,
the required condition is achieved by the time coordinate shift
$\delta t=\phi_1/\dot{\phi}_0$.
Then Hawking says that the change
of time coordinate will introduce inhomogeneous fluctuations
in the rate of expansion $H$. He and other authors take
(apparently, on the grounds of dimensionality only)
$\delta H\sim H^2\delta t$.
{}From here they come, implicitly or explicitly,
to the dimensionless amplitude of density perturbations
\begin{equation}
  {\delta\rho \over \rho} \sim {\delta H\over H}
\sim H\delta t \sim {H\phi_1 \over \dot{\phi}_0} \quad .
\end{equation}
Some authors write explicitly
$\phi_1 \sim H$
and $\delta\rho / \rho \sim H^2 / \dot{\phi}_0$.

The analysis has been done at the inflationary stage.
To obtain the today's amplitude of density perturbations
in wavelengths, say, of the order of the today's Hubble radius,
it is recommended to calculate the right hand side of Eq.~(65)
at the moments of time when the scales of our interest were
``crossing horizon'' during inflationary epoch. In one or another
version this formula appears in the most of inflationary literature
and because of numerous repetitions it has grown to the ``standard'' one.
According to this formula, the amplitude of density perturbations
becomes larger if one takes the $\dot{\phi}_0$ smaller.

The authors of [33-35] work with a specific scalar field potential,
so the numerical value of $\dot{\phi}_0$ and the numerical value
of $\delta\rho / \rho$ following from Eq.~(65) turn out
to be dependent on the self-coupling constant in the potential.
These authors are concerned about the unacceptably large amplitude
of density perturbations that they have produced.
But this is not a concern about the fact that the Einstein
equations play no role in this argumentation,
it is a specific detail in the scalar field potential that
the authors of~[33-35] do not like.

Now let us show the shortcomings of the argumentation in~[33-35].
Let us consider a scalar field $\phi$ with arbitrary potential.
Write the field as $\phi=\phi_0(t)+\phi_1(t)Q$ where $Q$
is the $n$-th spatial harmonic, $Q^{,i}_{,i}+n^2Q=0$.
Write the perturbed metric in the form
\[
ds^2 = -dt^2 + a^2(t)[(1+h(t)Q)\delta_{ij}
     + h_l(t)n^{-2}\, Q_{,i,j}]dx^i\, dx^j \quad .
\]
The de Sitter solution corresponds to $\dot{\phi}_0 =0$,
$a(t)\sim e^{Ht}$, and
$H(t)=\dot a /a={\rm const}$.
It follows from the Einstein equations that the (linear)
contribution $\epsilon_\phi$ of the scalar field perturbations
to the total energy density $\epsilon =\epsilon_0+\epsilon_\phi$
can be written as
\[
  \epsilon_\phi = \dot{\phi}_0 \left\{ \dot{\phi}_1-\phi_1
\big[ \ln(a^3\dot{\phi}_0)\big]^\cdot\right\} Q \quad .
\]
The contribution $\epsilon_\phi$, as well as other components
of the perturbed energy-momentum tensor, vanish in the de Sitter
limit $\dot{\phi}_0 \rightarrow 0$.

Thus, the first conclusion
we have to make is that in the de Sitter limit there is
no linear density perturbations at all. The scalar field
perturbations are uncoupled from gravity, they are not
accompanied by linear perturbations of the energy-momentum
tensor and they are not accompanied by linear perturbations
of the gravitational field. The general solution to the
perturbed Einstein equations is a set of purely coordinate
solutions which can be produced or totally removed by appropriate
coordinate transformations. The scalar field perturbations reduce
to a test field whose only role is to identify events in the
spacetime. One can still ask about a coordinate
system such that the surfaces of constant time $\tau$,
$\tau=\phi_1(t,{\bf x})$ are surfaces of constant $\phi$.
But the perturbation of the expansion rate of this new
coordinate system will have nothing to do with perturbations in
the energy density. Despite the presence of the test scalar
field, every space-like hypersurface is a surface of
constant energy density.

Now let us assume that $\dot{\phi}_0$ is not zero.
Transformation of time $\bar{t}=t+\chi (t)Q$ generates a Lie
transformation of the scalar field:
\[
  \phi_0(t)+\phi_1(t)Q \rightarrow \phi_0(t)
+ [\phi_1(t)-\dot{\phi}_0(t)\chi (t)]Q \quad .
\]
If one wants the transformed field to be homogeneous one takes
$\chi (t)=\phi_1(t)/\dot{\phi}_0(t)$. The same transformation
of time generates Lie transformations of the metric.
The transformed $g_{oo}$ component is
$\bar{g}_{oo} =-1+2\dot{\chi} Q$, the transformed $g_{ik}$
components are described by\\
$\bar{h}=h-2(\dot a /a)\chi$.
There appear also the $g_{oi}$ components but they will not
participate in our linear analysis. The expansion rate of
a given frame of reference is determined by the trace of
the deformation tensor~[36]:
\[
  D = {1\over 2\sqrt{-g_{oo}}}\,
{\partial (g_{ik}-g_{oi}g_{ok}/g_{oo})\over\partial t}\,
  g^{ik} \quad .
\]
In the linear approximation and before the transformation,
\[
  D \approx 3H + {1\over 2} (3\dot h - \dot h_l)Q \quad .
\]
After the transformation,
$\bar{D} = D - 3\dot H \chi Q$.

So, the introduced inhomogeneous fluctuation in the rate
of expansion is $\delta H = -\dot H \chi Q =\dot H \delta t$,
not $\delta H = H^2\delta t$ assumed in Refs.~[33-35].

The Einstein equation for energy density
$D^2/3 = \kappa\epsilon$ is satisfied before
and after the transformation, since the variation of $H$
is balanced by the variation of the energy density.
The transformed energy density is
\[
 \bar{\epsilon }
= \epsilon_0 + \epsilon_\phi - \dot\epsilon_0 \chi Q
= \epsilon_0 + \dot{\phi}_0^2
  \left( {\phi_1 \over \dot{\phi}_0}\right)^\cdot Q \quad .
\]

Thus, if one makes the $\dot{\phi}_0$ smaller, the energy
density perturbation decreases according to the Einstein
equations, and it increases according to the conjectures
of Refs.~[33-35]. The use of the correct expression
$\delta H \sim \dot H \delta t$ in Eq.~(65) makes the
$\delta\rho / \rho$ decreasing when the $\dot{\phi}_0$ is
decreasing.

The situation becomes even more disturbing if one recalls
that the formula (64) has been seemingly
confirmed and derived rigorously as a result of more
detailed studies. People did really write the perturbed
Einstein equations. Moreover, it was done in the framework
of the so-called gauge-invariant formalism, the whole purpose
of which is to eliminate coordinate solutions and to work
exclusively with something ``physical''. The basic mathematical
tool in these studies is the gauge-invariant potentials
$\Phi$ and $\Psi$ constructed from the components of the
perturbed metric.

As we have seen above, density perturbations in the scalar
field matter vanish
when $\dot{\phi}_0$ goes to zero, and there is no
density perturbations at all at the de Sitter stage.
This is true irrespective of the wavelength of the perturbation.
It can be shorter or much longer than the Hubble radius,
that is, it can be ``inside'' or ``outside'' the Hubble radius.
In particular, this is true of the perturbation whose
wavelength is such that it will grow by today to the scale of
today's Hubble radius.
The gauge-invariant potentials $\Phi$ and $\Psi$ are strictly
zero at the de Sitter stage. Why does then the today's amplitude
of the perturbation go to infinity, according to the ``standard''
formula, in the limit $\dot{\phi}_0 \rightarrow 0$ at the ``first
horizon crossing''? Because, the inflationary literature explains, the
amplitude will be almost infinitly enhanced in ``the course of
transition'' of the background equation of state from the
quasi-de Sitter one $p \approx - \epsilon$ to the radiation-dominated
$p={1\over 3}\epsilon$ or matter-dominated $p=0$ one.
The favorite concept in this argumentation is the ``constancy
of $\zeta$''. The often quoted papers are Ref.~[37], which uses
notations and equations of~[38], and Ref.~[39]
which summarizes the previous work and gives a
clearer exposition. We will follow equations and notations of
the paper~[39]. The advantage of this paper is that it contains
enough mathematical details to make it possible to follow
the spirit and the letter of calculations.

Mukhanov {\it et al.} [39] work with the gauge-invariant potentials
$\Phi$ and $\Psi$. According to one of the perturbed Einstein
equations, $\Phi = \Psi$.  In terms of the $\Phi$, the basic
equation of~[39] at the scalar field stage is
\begin{equation}
 \Phi^{\prime\prime} + 2
 {(a/\phi_0^\prime )^\prime \over (a/\phi_0^\prime )}
 \Phi^\prime - \nabla^2 \Phi + 2\phi_0^\prime
\left( { {\cal H} \over \phi_0^\prime}\right)^\prime
       \Phi = 0
\end{equation}
where ${}^\prime = d/d\eta$, $dt=ad\eta$, ${\cal H} = a^\prime /a$.
Equation~(66) is exactly the same equation as the basic
equation~(2.23) of Ref.~[37]. At the perfect fluid stage, the basic
equation of~[39] for ``adiabatic'' density perturbations is
\begin{equation}
  \Phi^{\prime\prime}
+ 3{\cal H}(1+c_s^2)\Phi^\prime - c_s^2 \nabla^2 \Phi
+ [2{\cal H}^\prime +(1+3c_s^2){\cal H}^2]\Phi = 0
\end{equation}
where $c_s^2 =\delta p/\delta\epsilon =\dot p_0/\dot\epsilon_0$.
Equations~(66) and (67) were derived
from the original perturbed Einstein equations with the help
of manipulations aimed at expressing the equations in terms
of the gauge-invariant potentials.

The authors of [39] introduce a new quantity $\zeta$ defined as
\begin{equation}
  \zeta \equiv {2\over 3}\, {H^{-1}\dot\Phi + \Phi \over 1+w}
        + \Phi \quad ,
\end{equation}
where $w=p_0/\epsilon_0$. This quantity is simply a new letter. As soon
as the function $\Phi$ is known, the function $\zeta$ can be calculated
from the definition (68). Using the definition of $\zeta$,
Eq.~(66) can be written in the form
\begin{equation}
   {3\over 2}\, \dot\zeta \, H(1+w)
= {1\over a^2} \nabla^2\Phi \quad ,
\end{equation}
and Eq.~(67) in the form
\begin{equation}
  {3\over 2} \dot\zeta H(1+w)
= {1\over a^2} c_s^2 \nabla^2\Phi \quad .
\end{equation}

Mukhanov {\it et al.} [39] consider perturbations with
wavelengths ``far outside the Hubble
radius for which $\nabla^2 \Phi$ can be neglected''. Neglecting the right
hand sides of Eqs.~(69), (70) they arrive, in this approximation,
at the equation $\dot\zeta \approx 0$ and the ``conservation law''
\begin{equation}
   \zeta \approx {\rm const} \quad .
\end{equation}

It is important to note that the derivation of this conservation law
did not require any knowledge about the initial data and solutions
for $\Phi$. The constant in the right hand side of Eq.~(71) emerges
as a universal number, irrespective of any particular solution
for $\Phi$. Although the ``constancy of $\zeta$'' is a favorite notion
in the inflationary literature, it appears that inflationists have never
asked what the origin and numerical value of this constant is.
If this constant is a universal number, is it equal to a billion,
or one, or zero? We will later show that this particular constant
must be equal to zero. However, without specifying the value of
this constant, it is often assumeed that it is not zero.

The next step in this argumentation proceeds as follows. Imagine
that the background equation of state changes during a short interval
of time from the initial $(i)$ $p \approx - \epsilon$ to the final
$(f)$ $p = {1\over 3}\epsilon$ or $p=0$, so that $1+w_i \ll 1$
while $1+w_f \approx 1$.  The authors of [39] consider the evolution
of a long-wavelength perturbation from the initial Hubble radius
crossing $(t_i)$ to the final Hubble radius crossing $(t_f)$.
They make additional assumptions, they assume that $\dot\Phi$
vanishes ``at very early and very late times''.
In the definition (68), they drop the term
with $\dot\Phi$ and simplify the quantity $\zeta$:
\[
   \zeta (t_i) \approx {5+3w(t_i)\over 3(1+w(t_i))} \Phi(t_i)
\]
\[
   \zeta (t_f) \approx {5+3w(t_f)\over 3(1+w(t_f))} \Phi(t_f)
\]
Then they refer to the constancy of $\zeta$,
$\zeta(t_i) = \zeta(t_f)$, and arrive at the formula
\begin{equation}
        \Phi(t_f) \approx {1+w(t_f)\over 1+w(t_i)}
        {5+3w(t_i)\over 5+3w(t_f)}\Phi(t_i)
\approx {1\over 1+w(t_i)}\Phi(t_i) \quad .
\end{equation}
(which, in fact, is in a conflict with the constatncy of $\zeta$, as we
will show later).

According to this formula, and since $1+w(t_i)$ can be arbitrarily
close to zero, the $\Phi(t_f)$ can be made arbitrarily large for
any nonvanishing $\Phi(t_i)$. Moreover, since the time of transition
from one equation of state to another can be arbitrarily short,
the tremendous increase of numerical value of the potential $\Phi$
is supposed to happen almost instantaneously. (The author of [32],
who reconfirmes the ``standard'' results, has even plotted a graph
for this jump.) The authors of Ref.~[39] emphasize that their
result~(72) (the actually published [39] formula (6.67) contains a
misprint: the position of symbols $w(t_i)$ and $w(t_f)$ should be
interchanged) is in a full agreement with previous studies and
Eq.~(64).  Formula (72) suggests an arbitrarily large production
of density perturbations for no other reason but simply because the
$1+w(t_i)$ was very close to zero. There is something strange with
this formula. [I realize well that what I qualify here as strange
is certainly considered by others as perfectly alright.
Otherwise somebody would raise a voice of protest against
the ease with which inflationists generate tremendous amounts
of various substances (some of the authors are even claiming that
they can ``overclose'' our Universe). However, judging from the
literature, it is not only that there are no voices of protest,
but there is rather an element of competition as for who was
the first to proclaim the ``standard'' inflationary results.
For instance, the authors of~[40] address the inflationary
claims about density perturbations as ``first quantitatively
calculated in~[33-35] [and which] have been successfully
quantitatively confirmed by the COBE discovery''.]

Let us now try to sort out what we are dealing with.

The first point to realize is that the basic equations (66),
(67) are, strictly speaking, incorrect. They are incorrect
in the following sense: they are constructed from a correct equation
and the first time derivative of the correct equation. Transforming
the original Einstein equations, the authors of~[38,39]
have effectively raised the order of differential equations.
If the correct equation is satisfied, the constructed equation
is satisfied too, but not {\it vise versa}. To get the feeling
of the danger involved, consider a correct equation $\dot x =0$
and a constructed equation $\ddot{x} - A\dot x = 0$ where $A$
is arbitrary constant. Solutions to the correct equation do satisfy
the constructed equation, but the latter one admits exponentially
growing solutions which are not allowed by the original
equation.

Let us start our analysis from Eq.~(66). Since the potential
$\Phi$ is an eigenfunction of the Laplace operator,
$\nabla^2\Phi =-n^2\Phi$,   we will replace
$\nabla^2\Phi$ with $-n^2\Phi$. We introduce also
a new function of the scale factor $a(\eta )$:
\[
\gamma = -\dot H/H^2 = 1+(a/a^\prime )^\prime
\]
and  use the background Einstein equations in order to express
the coefficients of Eq.~(66) in terms of the scale factor
$a(\eta )$ and its derivatives. Now, introduce a new variable
$\mu$ according to the definition
\begin{equation}
  \Phi = {1\over 2n^2}\, {a^\prime \over a} \gamma
  \left( {\mu\over a\sqrt\gamma}\right)^\prime
\end{equation}
So far, the variable $\mu$ is simply a new variable replacing
$\Phi$, but the importance of $\mu$ is in that the original
perturbed Einstein equations require this variable to satisfy
the equation
\begin{equation}
  \mu^{\prime\prime} + \mu
\left[ n^2 -
       {(a\sqrt\gamma )^{\prime\prime} \over a\sqrt\gamma}
\right] = 0 \quad .
\end{equation}
(For those interested in gauge-invariant potentials, I may
remark that $\mu$ is a genuine gauge-invariant variable
in the sense of~[38,39].)  With the help of Eq.~(73), Eq.~(66)
identically transforms to
\begin{equation}
\Bigg[ \mu^{\prime\prime} + \mu
\left[ n^2 - {(a\sqrt\gamma )^{\prime\prime}
       \over a\sqrt\gamma} \right] \Bigg]^\prime
- {(a\sqrt\gamma )^\prime \over a\sqrt\gamma}
\Bigg[ \mu^{\prime\prime} + \mu
\left[ n^2 - {(a\sqrt\gamma )^{\prime\prime}
       \over a\sqrt\gamma} \right] \Bigg] = 0 \, .
\end{equation}
If Eq. (74) is satisfied, Eq.~(75) is satisfied too,
but not {\it vice versa}. Equation~(75) is totally equivalent to
\begin{equation}
   {1\over a^2\gamma} \Bigg[ a^2\gamma
    \left( {\mu\over a\sqrt\gamma}\right)^\prime \Bigg]^\prime
   + n^2 {\mu\over a\sqrt\gamma} = X \quad ,
\end{equation}
where $X$ is arbitrary constant. Thus, Eq.~(66) involves an
arbitrary constant $X$ which, in fact, must be zero
according to Eq.~(74).

Let us now turn to the perfect fluid equation (67). The situation here
is similar to the scalar field case. Introduce a new variable $\nu$
according to the definition
\begin{equation}
  \Phi = {1\over 2n^2} {a^\prime \over a} {\gamma\over c_s^2}
\left( {\nu \over a\sqrt{\gamma/c_s^2}}\right)^\prime \quad .
\end{equation}
The importance of $\nu$ is in that the original perturbed Einstein
equations require this variable to satisfy the equation
\begin{equation}
  Z \equiv \nu^{\prime\prime}
+ \nu \Biggl[ n^2c_s^2 -
  {(a\sqrt{\gamma/c_s^2})^{\prime\prime}
    \over a\sqrt{\gamma/c_s^2}} \Biggr] = 0
\end{equation}
(The function $\nu$ is a genuine gauge-invariant variable.)
With the help of Eq.~(77), Eq.~(67) identically transforms to
\begin{equation}
    Z^\prime
  - {(a\sqrt{\gamma c_s^2})^\prime
    \over a\sqrt{\gamma c_s^2}} Z = 0
\end{equation}
This equation is totally equivalent to
\begin{equation}
   {1\over a^2\gamma}
\Biggl[ {a^2\gamma \over c_s^2}
\left( {\nu\over a\sqrt{\gamma/c_s^2}} \right)^\prime\Biggr]^\prime
     + n^2 {\nu\over a\sqrt{\gamma/c_s^2}} = Y
\end{equation}
where $Y$ is arbitrary constant. Thus, Eq.~(80) does also involve an
arbitrary constant $Y$ which, in fact, must be zero
according to Eq.~(78).  The role of these constants $X$ and $Y$
in the constancy of $\zeta$ argument we will discuss shortly.

Equations (74), (78) is all we need to solve, in order to find all the
perturbed components of the metric tensor and energy-momentum tensor.
There is no other way to find the evolution of density perturbations
except of doing the hard work of solving these equations,
imposing appropriate initial conditions, joing the solutions, etc.
(This is what is being done, convincingly or not, in Ref.~[3].
It is only great physicists can afford having one or several
the greatest mistakes of their life, we can not afford any.)
Among other things, these solutions will tell you what is the value
and time dependence of the functions $\zeta$ introduced by the
definitions (68), (73), (77). The correct equations do not show anything
like enormously large amplitude of today's density perturbations in
the limit of the de Sitter expansion. However, we need to return to
the concept of ``constancy of $\zeta$''  which pretends to answer
important physical questions without solving any equations at all.

Combine the definitions (68), (73) to show that $\zeta$ at the
scalar field stage is
\[
\zeta = {1\over 2n^2}\, {1\over a^2\gamma}
\Bigg[ a^2\gamma \left( {\mu\over a\sqrt\gamma}\right)^\prime
\Bigg]^\prime \quad .
\]
Combine the definitions (68), (77) to show that $\zeta$ at
the perfect fluid stage is
\[
       \zeta = {1\over 2n^2} {1\over a^2\gamma}
\Biggl[ {a^2\gamma\over c_s^2}
\left( {\nu\over a\sqrt{\gamma/c_s^2}} \right)^\prime \Biggr]^\prime
\]
The term with the Laplacian neglected in Eqs.~(66), (67) is the term
with $n^2$ in equations (76) and (80). Let us drop this term as the authors
of [39] do. Then, in this approximation, Eq.~(76) gives
$\zeta\approx X/2n^2 ={\rm const}$ and Eq.~(80)
gives $\zeta\approx Y/2n^2 ={\rm const}$. These relationships show that
$\zeta$ can be a universal constant, independent of any particular solution,
only if it is supported by the constant $X$ or, correspondigly, by
the constant $Y$. But, as we have shown above, these constants must be
equal to zero, if one is willing to work with correct equations.
The ``constancy of $\zeta$'' argument fails at the very beginning,
regardless of validity or not of the additional assumptions that have
been made on the route to Eq.~(72).
The conservation law $\zeta(t_i) = \zeta(t_f)$ degenerates to an empty
statement $0=0$, and nothing can be derived from it.

The idea of a free lunch based on the ``constancy of $\zeta$'' seems
to be so attractive that the following question is often being asked.
Suppose that not for all solutions for $\Phi$, but for some of them,
suppose that not in the leading order $n^{-2}$, but in the next
order, suppose, nevertheless, that $\zeta$ calculated from this
specially chosen solution for $\Phi$ is approximately a nonzero
constant (which is indeed possible). Why cannot we return to Eq.~(68)
and repeat all the arguments that have seemingly led us
from the definition (68) to the result (72)?
For instance, the authors of [31] have even considered,
along this line, a concrete model consisting of two consecutive
power-law scale factors. They speak about an initially small potential
$\Phi$ being ``distributed'' after the transition point ``into two modes,
both very large in amplitude, one which decays, the other yielding''
the ``standard'' result (72).  And this is how, they argue, the
``scalar perturbations can be very strongly amplified in the course
of the transition''.

So, we also need to consider this model and explore what is contained
in the definition (68) treated as an equation for $\Phi$.

First of all, use the background equation
\[
  1+w = -{2\over 3}{\dot H\over H^2}
\]
and write $\zeta$ in a more convenient form:
\begin{equation}
   \zeta = -{H^2 \over a\dot H}
\left( {a\over H}\Phi\right)^{\cdot}\quad .
\end{equation}
Integrate this equation to produce the general solution
\begin{equation}
  \Phi = {H\over a}
   \left[ C+\int a\zeta \left( {1\over H}\right)^{\cdot} dt \right]
\end{equation}
where $C$ is arbitrary integration constant. So far, $\zeta$ can be
an arbitrary function. Now assume that $\zeta$ is a constant,
$\zeta = \zeta_0 = {\rm const}$, and write
\begin{equation}
  \Phi = \zeta_0
\left[ 1-{H\over a} \int_{t_i}^t adt \right] + C{H\over a} \quad .
\end{equation}
[The constants $\zeta_0$ and $C$ can be related to the
coefficients in front of two linearly  independent solutions to
Eq.~(74).  In the long wavelength approximation,
\[
  {\mu \over a\sqrt\gamma} = C_1
\left[ 1-n^2 \int{1\over a^2\gamma}
\left( \int a^2\gamma d\eta \right) d\eta \right]
     + C_2 \int {d\eta \over a^2\gamma} + \ldots \quad .
\]
Using the definition (73) one can show that
$\zeta_0 =-{1\over 2}C_1$, $C={C_2\over 2n^2}$.]

As the second step, let us consider, together with Deruelle and
Mukhanov [31], an expanding model which decribes a transition from
one power-law scale factor to another power-law scale factor.
Let us write
\begin{equation}
  a_i = a_1t^{p_1} \quad , \qquad
  a_f = a_2(t-t_\ast )^{p_2} \quad .
\end{equation}
The scale factor and its first time derivative are continuous functions
at the transition point $t = t_1$. This requires
\[
  a_2=a_1\left( {p_1\over p_2}\right)^{p_2} t_1^{p_1-p_2} \quad , \qquad
  t_\ast = t_1\left( 1- {p_2\over p_1}\right) \quad .
\]
To be closer to the inflationary results, one can keep in mind
that $p_1 \gg 1$, while $p_2$ is 1/2 or 2/3.

Now we will follow the time evolution of $\Phi$ specifically in this
model. Let $t_i$ be the time when our perturbation first
crosses the Hubble radius: $a(t_i)H(t_i)=n$.
Let $t_f$ be the time when the perurbation returns back inside the
Hubble radius: $a(t_f)H(t_f)=n$.  The functions $a(t)$ and $H(t)$
are continuous all the way from $t_i$ to $t_f$, and so is the
function (83). The initial value of the potential is
\begin{equation}
  \Phi(t_i) =\zeta_0 + C{H(t_i)\over a(t_i)} \quad ,
\end{equation}
the final value is
\begin{equation}
  \Phi (t_f) = \zeta_0 I+C{H(t_f)\over a(t_f)}
\end{equation}
where
\[
  I \equiv 1 - {H(t_f)\over a(t_f)} \int_{t_i}^{t_f} adt \quad .
\]
Combining (85) and (86) we find
\begin{equation}
  \Phi(t_f) = \Phi(t_i)I + C
\left[ {H(t_f)\over a(t_f)} - {H(t_i)\over a(t_i)} I\right] \quad .
\end{equation}
The integral $I$ can be easily calculated for the scale factor (84).
Since $a(t_f) \gg a(t_i)$, $p_1 \gg p_2$, and $a(t_1)H(t_1) \gg n$,
we have approximately
\[
  I \approx {1\over p_2+1} \quad .
\]
Formula (87) reduces to
\[
   \Phi(t_f) = {1\over p_2+1} \Phi(t_i)
- C{H(t_i)\over a(t_i)} {1\over p_2+1} \quad .
\]
The constant $C$ could be set to zero from the very beginning.
In any case, the term $C(H(t_i)/a(t_i))$ is of the same order
or smaller than $\Phi(t_i)$. By setting $C=0$, we arrive at
\[
 \Phi(t_f) \approx {1\over p_2+1} \Phi(t_i) \quad .
\]
There is nothing like tremendous jumps of $\Phi$ at the
transition point.  The ``constancy of $\zeta$'' effectively
translated into constancy of $\Phi$.  [The fact that $\Phi$ is
approximately constant for wavelengths longer than the Hubble
radius follows also from the exact solution to Eq.~(74) which can
be found in case of power-law scale factors.]

The best what we can say about the ``standard'' inflationary formula
is that it does not follow from correct equations.

It is important to recall that the ``standard'' inflationary results
seem to be an indispensable tool in cosmology of our days (and possibly
in the next Millennium too, see for example [41,10]).
The expected relative contributions of density perturbations and
gravitational waves to the observed microwave background anisotropies
are a subject of active study. In the center of discussion are
usually the ``consistency relations'' which state that the
ratio of density to gravity-wave
contributions goes to infinity when the spectrum of perturbations
approaches the most favorite, Harrison-Zeldovich form. In reality,
as we have shown above, these ``consistency relations''
are simply a manifistation of inconsistency of the ``standard''
inflationary theory from which they are derived.

In conclusion, if the ``standard'' inflationary results are
incorrect and cannot be trusted, what is the amount of density
perturbations that can be generated in the early Universe?
My part of answer is formulated in Ref.~[3,7].

\acknowledgments

I appreciate discussions with Yu.\ V. Sidorov on squeezed states,
and collaboration with J. Martin on cosmological
perturbations. This work was supported by NASA grants NAGW~2902, 3874
and NSF grant 92-22902.


\newpage
\begin{references}
\bibitem{1}
C. J. Isham, Structural Issues in Quantum Gravity,
gr-qc 9510063.
\bibitem{2}
G. F. Smoot, {\it et al.,} Astrophys. J. {\bf 396}, L1 (1992).
\bibitem{3}
L. P. Grishchuk, Phys. Rev. D {\bf 50}, 7154 (1994).
\bibitem{4}
L. P. Grishchuk and Y. V. Sidorov, Phys. Rev. D {\bf 42}, 3413 (1990);
Class. Quantum Grav. {\bf 6}, L161 (1989).
\bibitem{5}
Special issues: J. Mod. Opt. {\bf 34} (6,7) (1987);
J. Opt. Soc. Am. {\bf 84}, (10) (1987); NASA Conference Publication
3135 (1992).
\bibitem{6}
Special issue: Physica Scripta (Eds. W. P. Schleich and S. M. Barnett)
{\bf T 48} (1993).
\bibitem{7}
L. P. Grishchuk. Cosmological Perturbations of
Quantum-Mechanical
Origin and Anisotropy of the Microwave Background Radiation.
In: {\it Current Topics in Astrofundamental Physics:
The Early Universe} (Eds. N. Sanchez and A. Zichichi)
NATO ASI Series C, Vol. 467, p. 205 (Kluwer Academic
Publishers, 1995).
\bibitem{8}
L. P. Grishchuk. Comment on ``Cosmological Perturbations of a Relativistic
Condensate''. Report WUGRAV-95-12, gr-qc~9506010.
\bibitem{9}
J. M. Niedra and A. I. Janis, G.R.G. {\bf 15}, 241 (1983).
\bibitem{10}
P. J. Steinhardt. Cosmology at the Crossroads. Invited talk at the
Snowmass meeting (to be published), astro-ph 9502024.
\bibitem{11}
R. K. Sachs and A. M. Wolfe, Astrophys. J. {\bf 1}, 73 (1967).
\bibitem{12}
L. P. Grishchuk, Phys. Rev. D {\bf 48}, 3513 (1993).
\bibitem{13}
L. P. Grishchuk, Phys. Rev. D {\bf 48}, 5581 (1993).
\bibitem{14}
P. Paradoksov, Uspekhi Fiz. Nauk, {\bf 89}, 707 (1966)
[Sov. Phys. Uspekhi, {\bf 9}, 618 (1967)].
\bibitem{15}
G. A. Korn and T. M. Korn, {\it Mathematical Handbook
for Scientists and Engineers} (McGraw-Hill Book Co., 1968).
\bibitem{16}
M. Fisz, {\it Probability Theory and Mathematical Statistics}
(J. Wiley \& Sons, 1967).
\bibitem{17}
I. S. Gradshteyn and I. M. Ryzhik, {\it Table of Integrals,
Series, and Products} (Academic Press, 1980).
\bibitem{18}
M. Abramowitz and I. A. Stegun, (eds), {\it Handbook of Mathematical
Functions} (Dover Publishing, 1972).
\bibitem{19}
C. L. Bennett, {\it et al.,} Astrophys. J. {\bf 436}, 423 (1994).
\bibitem{20}
E. L. Wright, {\it et al.,} Astrophys. J. {\bf 436}, 443 (1994).
\bibitem{21}
For gravitational waves this formula is in fact known for long
time.  See Eq.~(5b) in L.~P.~Grishchuk, Ann. N.Y. Acad. Sci.
{\bf 302}, 439 (1977) where the notational replacement
$h(n)=A(n)$ should be made and where $A(n)$ should be interpreted
as the {\it final} (generated) {\it characteristic} amplitude,
$A^2(n)\sim B^2(n)n^{2(1+\beta )}$, while $B(n)$ should be
interpreted as the {\it initial} (zero-point quantum)
{\it characteristic} amplitude, $B(n)\sim n$, in accord
with the scale factor, Eq.~(15) above, in which the $\eta$-time
grows from $-\infty$.
\bibitem{22}
M. White, D. Scott, and J. Silk, Annu. Rev. Astron. Astrophys.
{\bf 32}, 319 (1994).
\bibitem{23}
L. F. Abbott and M. B. Wise, Astrophys. J. {\bf 282}, L47 (1984).
\bibitem{24}
J. R. Bond and G. Efstathiou, M.N.R.A.S. {\bf 226}, 655 (1987).
\bibitem{25}
R. Fabbri, F. Lucchin and S. Matarrese, Astrophys. J. {\bf 315},
1 (1987).
\bibitem{26}
R. Scaramella and N. Vittorio, Astrophys. J. {\bf 353}, 372 (1990).
\bibitem{27}
L. Cayon, E. Martinez-Gonzalez and J. L. Sanz, M.N.R.A.S. {\bf 253},
599 (1991).
\bibitem{28}
M. White, L. M. Krauss and J. Silk, Astrophys. J. {\bf 418}, 535 (1993).
\bibitem{29}
R. Scaramella and N. Vittorio, Astrophys. J. {\bf 411}, 1 (1993).
\bibitem{30}
A. M. Yaglom, {\it An Introduction to the Theory of Stationary
Random Functions} (Prentice-Hall Inc., 1962).
\bibitem{31}
N. Deruelle and V. F. Mukhanov, ``On Matching Conditions for
Cosmological Perturbation'', preprint gr-qc 9503050.
\bibitem{32}
R. R. Caldwell, ``On the Evolution of Scalar Metric Perturbations
in an Inflationary Cosmology'', preprint gr-qc 9509027.
\bibitem{33}
S. W. Hawking, Phys. Lett. {\bf 115B} 295, 1982).
\bibitem{34}
A. A. Starobinsky, Phys. Lett. {\bf 117B}, 175 (1982).
\bibitem{35}
A.H. Guth and S.-Y. Pi, Phys. Rev. Lett. {\bf 49}, 1110 (1982).
\bibitem{36}
A. L. Zel'manov, Doklady Acad. Nauk USSR {\bf 107}, 815 (1956)
[Sov. Phys. Doklady {\bf 1}, 227 (1956)].
\bibitem{37}
J. M. Bardeen, P. J. Steinhardt, and M. S. Turner, Phys. Rev. D
{\bf 28}, 679 (1983).
\bibitem{38}
J. M. Bardeen, Phys. Rev. D {\bf 22}, 1882 (1980).
\bibitem{39}
V. F. Mukhanov, H. A. Feldman, and R. H. Brandenberger,
Phys. Reports {\bf 215}, 6 (1992).
\bibitem{40}
D. Polarski and A. A. Starobinsky, ``Semiclassicality and
Decoherence of Cosmological Perturbations'', Report LMPM 95-4, gr-qc 9504030.
\bibitem{41}
J. A. Frieman, ``Cosmological Models at the Millennium'',
FERMILAB-Conf-95/151-A, 1995.
\bibitem{42}
L. P. Grishchuk, ``Statistics of the Microwave Background
Anisotropics Caused by the Squeezed Cosmological Perturbation'',
Report WUGRAV-95-6, gr-qc~9504045
\end{references}
\end{document}